\documentclass[twocolumn,numberedappendix]{emulateapj}
\shorttitle{PAPER-32 Power Spectrum Results }
\shortauthors{Parsons, et al.}

\usepackage{amsmath}
\usepackage{graphicx}
\usepackage[figuresright]{rotating}
\usepackage{natbib}
\newcommand{\BigO}[1]{\mathcal{O}(#1)}

\newcommand{\F}{\mathbf{F}}
\newcommand{\M}{\mathbf{M}}
\newcommand{\Q}{\mathbf{Q}}
\newcommand{\C}{\mathbf{C}}
\newcommand{\N}{\mathbf{N}}
\newcommand{\G}{\mathbf{G}}
\newcommand{\beq}{\begin{equation}}
\newcommand{\eeq}{\end{equation}}

\newcommand{\x}{\mathbf{x}}
\def\k{\mathbf{k}}
\def\r{\mathbf{r}}
\def\q{\mathbf{q}}
\def\b{\mathbf{b}}
\def\kp{\mathbf{k}^\prime}
\def\kpp{\mathbf{k}^{\prime\prime}}
\def\V{\mathbb{V}}
\def\At{\tilde{A}}
\def\Vt{\tilde{V}}
\def\Tt{\tilde{T}}
\def\tb{\langle T_b\rangle}

\begin{document}
\title{New Limits on 21cm EoR From PAPER-32 Consistent with an X-Ray Heated IGM at \lowercase{$z=7.7$}}
%\title{PAPER-32: A 21cm Reionization Upper Limit That Favors X-Ray Heating}

\author{
Aaron R. Parsons\altaffilmark{1,2},
Adrian Liu\altaffilmark{1},
James E. Aguirre\altaffilmark{3},
Zaki S. Ali\altaffilmark{1},
Richard F. Bradley\altaffilmark{4,5,6},
Chris L.  Carilli\altaffilmark{7},
David R. DeBoer\altaffilmark{2},
Matthew R. Dexter\altaffilmark{2},
Nicole E. Gugliucci\altaffilmark{6},
Daniel C. Jacobs\altaffilmark{8},
Pat Klima\altaffilmark{5},
David H. E. MacMahon\altaffilmark{2},
Jason R. Manley\altaffilmark{9},
David F. Moore\altaffilmark{3},
Jonathan C. Pober\altaffilmark{1},
Irina I. Stefan\altaffilmark{10},
William P. Walbrugh\altaffilmark{9}}

\altaffiltext{1}{Astronomy Dept., U. California, Berkeley, CA}
\altaffiltext{2}{Radio Astronomy Lab., U. California, Berkeley, CA}
\altaffiltext{3}{Dept. of Physics and Astronomy, U. Pennsylvania, Philadelphia, PA}
\altaffiltext{4}{Dept. of Electrical and Computer Engineering, U. Virginia, Charlottesville, VA}
\altaffiltext{5}{National Radio Astronomy Obs., Charlottesville, VA}
\altaffiltext{6}{Dept. of Astronomy, U. Virginia, Charlottesville, VA}
\altaffiltext{7}{National Radio Astronomy Obs., Socorro, NM}
\altaffiltext{8}{School of Earth and Space Exploration, Arizona State U., Tempe, AZ}
\altaffiltext{9}{Square Kilometer Array, South Africa Project, Cape Town, South Africa}
\altaffiltext{10}{Cavendish Lab., Cambridge, UK}

\begin{abstract}
We present new constraints on the 21cm Epoch of Reionization (EoR) power spectrum derived from 3 months of observing
with a 32-antenna, dual-polarization deployment of
the Donald C. Backer Precision Array for Probing the Epoch of Reionization (PAPER) in
South Africa.  In this paper, we demonstrate the efficacy of the delay-spectrum approach to
avoiding foregrounds, achieving over 8 orders of magnitude of foreground
suppression (in mK$^2$).  Combining this approach with a procedure for removing off-diagonal covariances arising from instrumental systematics, we achieve
a best $2\sigma$ upper limit of $(41 {\rm mK})^2$ for
$k=0.27\ h\ {\rm Mpc}^{-1}$ at $z=7.7$.  
This limit falls within an order of magnitude of the brighter predictions of the expected 21cm 
EoR signal level.  Using the upper limits set by these measurements, we generate new constraints
on the brightness temperature of 21cm emission in neutral regions for various reionization models.  We show
that for several ionization scenarios, our measurements are inconsistent with cold reionization.
That is, heating of the neutral intergalactic medium (IGM) is
necessary to remain consistent with the constraints we report.
Hence, we have suggestive evidence that
by $z=7.7$, the HI has been warmed 
from its cold primordial state, probably by X-rays from high-mass X-ray binaries or mini-quasars.
The strength of this evidence depends on the ionization state of the IGM, which we are
not yet able to constrain.  
This result is consistent with standard predictions for how reionization might have proceeded.

%we also present new measurements of the spectrum of Pictor A from 100 to 200 MHz, substantially
%revising older measurements presented in the literature.
\end{abstract}

\section{Introduction}

The Donald C. Backer Precision Array for Probing the Epoch of Reionization
(PAPER; \citealt{parsons_et_al2010})\footnote{\url{http://eor.berkeley.edu/}},
a dedicated experiment
employing non-tracking, dual-polarization dipole antennas in a 100--200-MHz band,
is one of several
radio telescopes aiming to 
measure the power spectrum of highly redshifted 21cm emission to inform our understanding
of cosmic reionization \citep{furlanetto_et_al2006,morales_wyithe2010,pritchard_loeb2012}. 
Other such telescopes include
the Giant Metre-wave Radio Telescope (GMRT;
\citealt{paciga_et_al2013})\footnote{\url{http://gmrt.ncra.tifr.res.in/}},
the LOw Frequency ARray (LOFAR; \citealt{yatawatta_et_al2013})\footnote{\url{http://www.lofar.org/}},
and the Murchison Widefield Array (MWA; \citealt{bowman_et_al2012}; \citealt{tingay_et_al2013})\footnote{\url{http://www.mwatelescope.org/}}.
PAPER consists of two distinct arrays: a 32-antenna deployment at the NRAO site 
near Green Bank, WV,
which is used primarily for engineering investigations and field testing, and a 64-antenna 
deployment in the Square Kilometre Array South Africa (SKA-SA) reserve in the Karoo desert near Carnarvon.
PAPER South Africa
(PSA) is used primarily for science observations, and provides the data 
on which this paper is based.

Recent advances in our understanding of how smooth-spectrum foreground emission can be isolated
from the 21cm EoR signature via delay-spectrum analysis (\citealt{parsons_et_al2012b}; hereafter P12b), and how 
maximum-redundancy array configurations can substantially improve the sensitivity of
power spectrum measurements in the low signal-to-noise regime 
(\citealt{parsons_et_al2012a}; hereafter P12a), have resulted in dramatic improvements in PAPER's prospects for
constraining the power spectrum of 21cm reionization.
In this paper, we provide a first look at the application
of delay-spectrum analysis to maximum-redundancy PAPER observations, culminating in a new upper limit on the
power spectrum of 21cm emission from reionization.

This paper is structured as follows.
In \S\ref{sec:observations} we describe the observations used,
\S\ref{sec:analysis} details the calibration and analysis pipeline,
in \S\ref{sec:results} we describe the results of the analysis and the
constraints we are able to place on reionization, and
we conclude in \S\ref{sec:conclusion} with 
a discussion of the efficacy of delay-spectrum analysis, as well as
PAPER's near-term prospects for improving upon these results.

\section{Observations}
\label{sec:observations}

\begin{figure*}\centering
\includegraphics[width=\columnwidth]{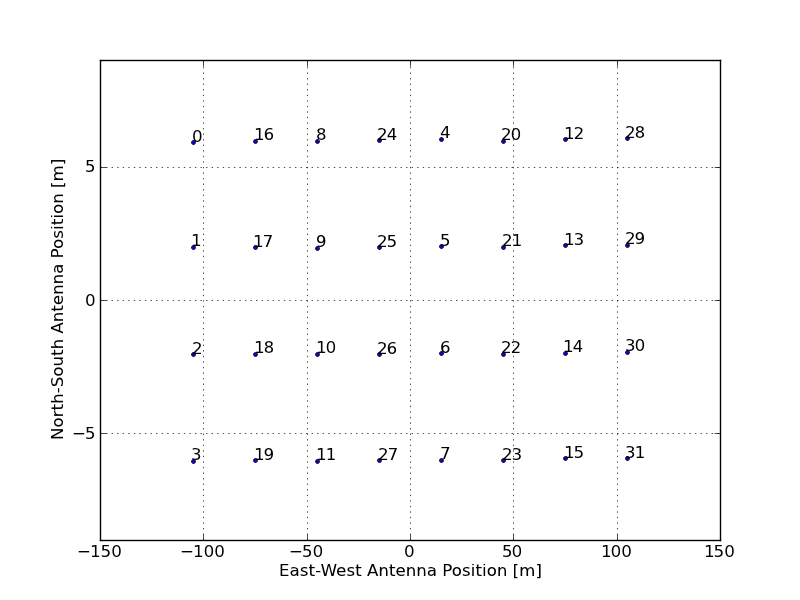}
\includegraphics[width=\columnwidth]{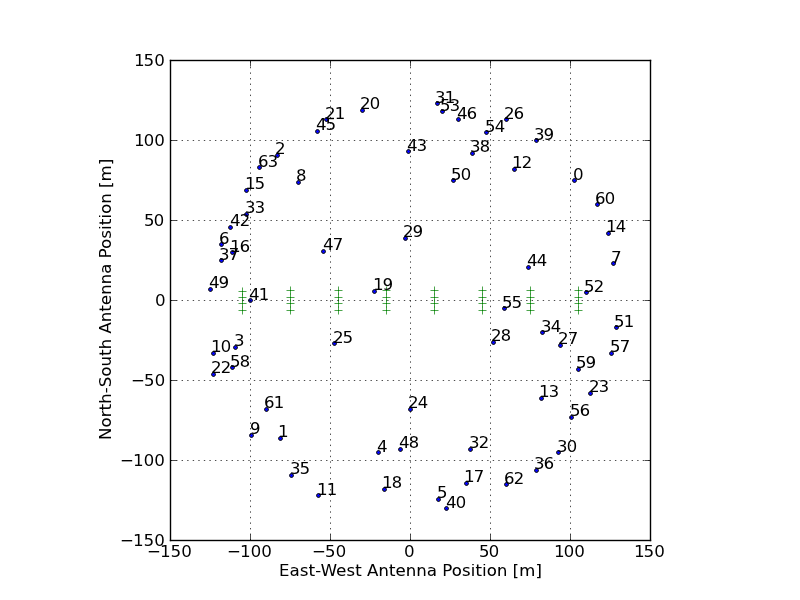}
\caption{Left: The 32-antenna, dual-polarization, maximum-redundancy array configuration used by the PAPER deployment
in the Karoo Radio Observatory site in South Africa for power-spectrum observations.  
For power-spectrum analysis, we use the baselines connecting antennas 
between adjacent columns for north-south deflections of zero (0-16, 1-17, 2-18, 3-19, 8-16, 9-17, etc.),
one (1-16, 2-17, etc.), and minus one (0-17, 1-18, etc.).
Note the expanded scale in the vertical axis.  
Right: The 64-antenna, single-polarization, minimum-redundancy array configuration on the same site,
whose imaging observations were used to characterize the spectrum of Pictor A in \citet{jacobs_et_al2013}.
The maximum redundancy positions are shown in green for comparison.
}\label{fig:antpos}
\end{figure*}

In this section, we summarize the salient features of the observations used in our analysis.  For more
details regarding the PAPER telescope, its drift-scan observing mode, and primary beam shape,
we refer the reader to \citet{parsons_et_al2010}, \citet{jacobs_et_al2011}, \citet{pober_et_al2012},
and \citet{stefan_et_al2013}.

Our analysis centers on
the XX and YY linear polarization products of 32 PAPER antennas 
deployed in the maximum-redundancy configuration shown in the left panel of Figure \ref{fig:antpos}.
All baselines within this core were used in the calibration steps described in \S\ref{sec:analysis}.
For this power-spectrum analysis, we use only a subset of the baselines: the 28 antenna pairs connecting
adjacent north-south columns in the strictly east-west direction (0-16, 1-17, 2-18, 3-19, 8-16, etc.),
as well as the 21 slanted just north of east (1-16, 2-17, 3-18, 8-17, etc.), and the 21 just south of
east (0-17, 1-18, 2-19, 9-16, etc.).  Within each of these groups, baselines
are all instantaneously redundant; they sample the same Fourier modes of the sky.  
Taken together, these 28+21+21=70 baselines represent
the majority of the total instantaneous power-spectrum 
sensitivity of the 32-antenna, maximum-redundancy observations (P12a).

For each baseline, a 100-MHz band from 100 to 200 MHz was
divided into 2048 frequency channels, integrated for 10.7 seconds, and recorded.
%Observations spanned a 59-day period from Dec. 7, 2011 to Feb. 4, 2012, corresponding to
%JD2455903.0 to JD2455962.2.  In this period, observations were recorded for 55 days.
%Over this period, we selected for analysis observations in the LST range of 1:30 to 6:30, corresponding to 
Observations spanned a 103-day period from Dec. 7, 2011 to Feb. 4, 2012, corresponding to
JD2455903 to JD2456006.  In this period, observations were recorded for 92 days.
Over this period, we selected for analysis observations in the LST range of 1:00 to 12:00, corresponding to 
a colder patch in the synchrotron sky that dominates the system temperature.
%In summary, the dataset on which this analysis is based consists of 55 full days observed with 28+21+21 east-west
In summary, the dataset on which this analysis is based consists of 92 full days observed with 28+21+21 east-west
oriented baselines of length 30 m, covering a window of 11 sidereal hours.

For the purpose of characterizing the flux density of Pictor A to establish a flux scale for the
power spectrum measurements, we also made use of data observed from JD2455747.1 to JD2455748.1
from a 64-antenna single (XX) polarization deployment of the 
PAPER array in a minimum redundancy configuration (see Figure \ref{fig:antpos}, right panel)
more suited for imaging.  The details
of these observations and their calibration are given in \citet{jacobs_et_al2013} and 
\citet{pober_et_al2013}.  Observation bandwidth,
frequency resolution, and integration time match the maximum-redundancy
observations reported above.

\section{Calibration and Analysis}
\label{sec:analysis}

\begin{figure*}\centering
\includegraphics[width=1.85\columnwidth]{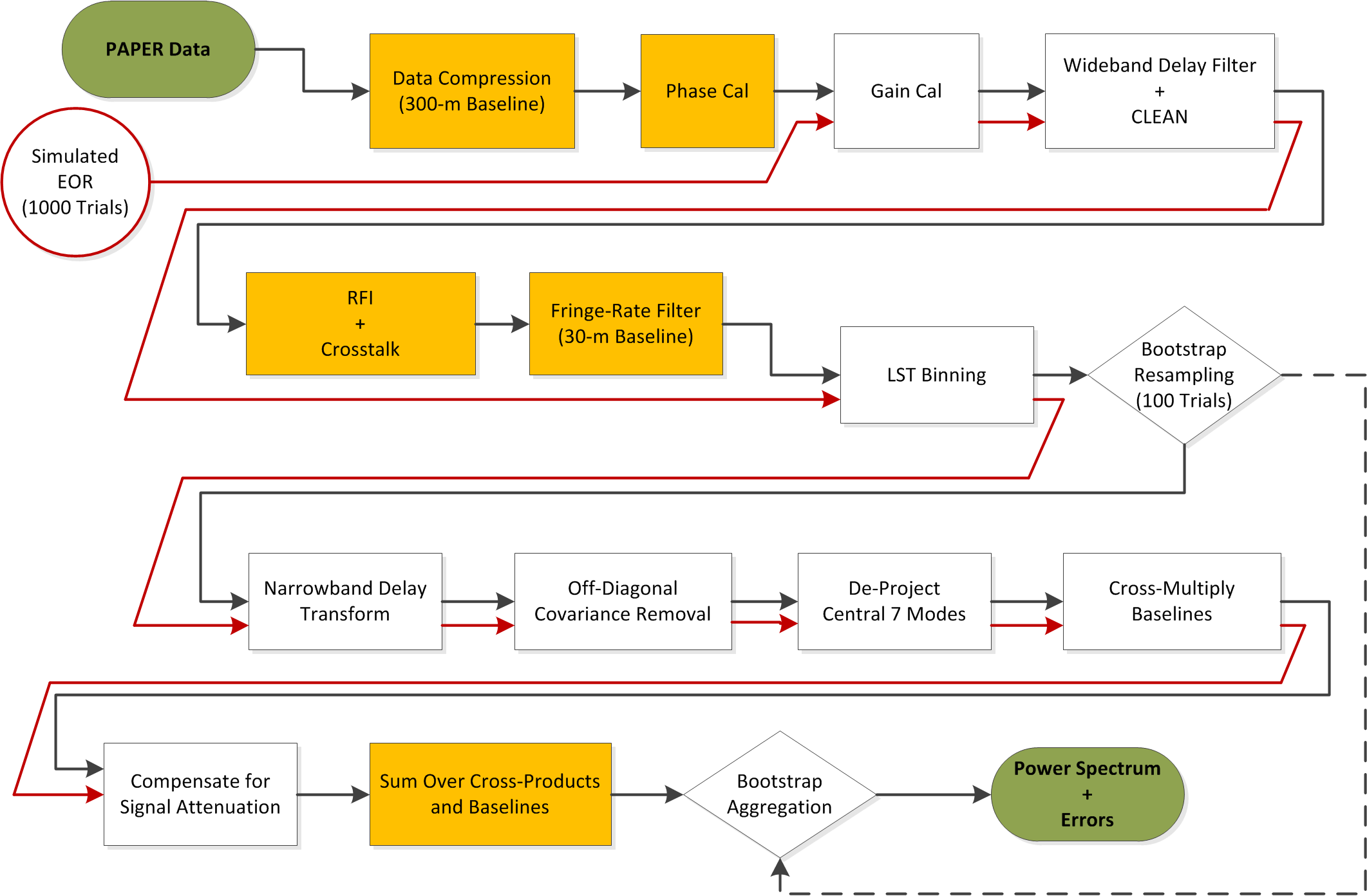}
\caption{Data flow for the power-spectrum analysis described in \S\ref{sec:analysis}.  Solid black lines
indicate data flow.  Solid red lines indicate Monte Carlo simulations used to assess signal
attenuation resulting from the analysis.  
Dotted lines indicate information flow for calculating
errors via bootstrapping.
Yellow indicates analysis steps not included in simulations
for which signal loss is calculated to be negligibly small.  
}\label{fig:data_flow}
\end{figure*}

All analysis described in this section were implemented using the Astronomical Interferometry in PYthon
(AIPY)\footnote{\url{http://pypi.python.org/pypi/aipy}} software toolkit.  This package is under revision 
control\footnote{\url{http://github.com/AaronParsons/aipy}}; our analysis is based on the version of AIPY
committed under the hash tag dff2f4dba731240ced5cc883895a80cf714fc12e.  An overview of the analysis in
this section is illustrated in Figure \ref{fig:data_flow}.

\subsection{Pre-processing}
\label{sec:preprocessing}

%xrfi_simple.py -a 1 --combine -t 20 -c 0_130,755_777,1540,1704,1827,1868,1885_2047 --df=6 
%ddr_filter_coarse.py -a 1 -p xx,yy --clean=1e-3 --maxbl=300 --output=ddr --invert 
%xrfi_simple.py -a 1 --combine -t 20 -n 4 ${FILE}E --to_npz=${FILE}E.npz
%xrfi_simple.py -a all --combine -t 20 ${FILE} --from_npz=${FILE}E.npz
%ddr_filter_coarse.py -a all --clean=1e-4 --maxbl=300 

In pre-processing, we use delay/delay-rate (DDR) filters \citep{parsons_backer2009} to help identify 
radio-frequency interference (RFI) events, and as part of a data compression
technique that reduces data volume by over a factor of 40.
Further details are supplied in Appendix \ref{app:data_compression}.
%First, we remove known RFI transmission bands and analog filter band edges, and then 
%coarsely flag data at 6$\sigma$ to remove RFI events after performing a derivative in the frequency direction
%to suppress celestial emission.
%Next, we suppress foreground emission by applying a DDR filter to remove delays and delay-rates 
%within the horizon limits for a 300-m baseline (the maximum length of any PAPER baseline).
%We derive an RFI mask by flagging 4$\sigma$ events in the residuals, and we apply these flags back
%to the unfiltered data.  Finally, we compress the data by applying a 
%DDR filter that preserves emission within the horizon limit of a 300-m baseline, deconvolving to suppress 
%flagging artifacts, and down-sampling the result in time/frequency domain to the critical Nyquist rate of
%the DDR filter.
The result is that the 2048 original frequency channels become 203, and the
60 original time samples per 10-minute file become 14, for an effective integration time
of 43s.  Our chosen output integration time is slightly shorter than
the 45.7 s interval dictated by the maximum fringe rate of a 300-m baseline.
These data contain all emission that
rotates with the sky and falls in the delay range
$|\tau|<1000$ ns, corresponding to $-0.5 < k_\parallel < 0.5\ h\ {\rm Mpc}^{-1}$ at
$z=7.7$ (164 MHz), but compressed in data volume by over a factor of 40.

\subsection{Phase Calibration}

Phase calibration is achieved using standard interferometric self-calibration with the addition of
constraints derived from the substantial instantaneous redundancy of the array
\citep{wieringa1992,liu_et_al2010}.
GPS surveying and total-station laser ranging located all antennas to within
$\pm$10 cm from a perfect grid, based on agreement between the two surveying methods.
These antenna coordinates are used to calibrate the electrical delay of each
antenna signal path on the basis of redundancy.  Because of the high degree of instantaneous redundancy
in this array, we are able to derive per-antenna delays for each signal path relative to only two 
parameters that could not be solved for on the basis of redundancy: the relative electrical delay
between the antennas in fiducial east-west and north-south baselines (0-16 and 0-1, respectively).
We find per-antenna delays averaged over each day of observation to agree
within $\pm$0.1 ns between days.  
Since this residual delay is a factor of 1000 smaller than the 100-ns width of the delay bins
used in the power spectrum analysis in \S\ref{sec:dspec_crossmult}, the signal attenuation that
would result from averaging over the phase variation associated with this residual delay is 
very small (of order $10^{-6}$).
On this basis, we adopt a single per-antenna delay solution 
for the entire 92-day observation.
These per-antenna delay solutions for all baselines are incorporated into a global
self-calibration step that derives the remaining two unknown phase parameters using
the central components of Centaurus A, Fornax A, and Pictor A.

The use of redundancy in calibration allows antenna-based gain and phase
parameters to be derived from the ratios of baselines within instantaneously
redundant sets.  Because these instantaneously redundant baselines measure the
same sky signal, the signal with each set of redundant baselines essentially
becomes a single degree of freedom that, by taking ratios, is projected out
prior to calibration.  Hence, there is no possibility of signal loss associated
with the redundant aspect of calibration.

\subsection{Flux Calibration}

\subsubsection{An Absolute Flux Scale from Pictor A}
\label{sec:pic_spec}

%\begin{figure*}\centering
%%\includegraphics[width=0.8\columnwidth]{plots/data/pic_spec_cal/pic_spec.png}
%\includegraphics[width=0.8\columnwidth]{plots/data/pic_spec_cal/pic_thumb-spec.png}
%\caption{Measured spectrum of Pictor A (upper black) and 2331-416 (lower black), with vertical bars indicating 
%2$\sigma$ confidence intervals for PAPER measurements binned in 10-MHz intervals.
%Solid black lines indicate the best-fit power law over the PAPER band for each of these sources.
%The spectrum of 2331-416 was calibrated to a power law with $S_{150}=31.9\ {\rm Jy}$, $\alpha=-0.69$
%(dotted cyan), based on a fit to measurements in the literature (cyan).
%Using this calibration,
%the best-fit power law to the spectrum of Pictor A measured by
%PAPER is $S_{150}=388.1\ {\rm Jy}\pm6.2$, $\alpha=-0.87\pm 0.12$.  Including measurements at
%other frequencies in the literature (magenta),
%the best-fit power law (dotted magenta)
%becomes $S_{150}=389.2\ {\rm Jy}\pm5.8$, with $\alpha=-0.76\pm0.01$.  This is the value to
%which the power spectrum measurements (see Figure \ref{fig:flux_cal}) are calibrated.
%Measurements at frequencies below 1 GHz drew on data in 
%\citet{slee1995}, \citet{kuhr_et_al1981}, \citet{large_et_al1981}, and \citet{burgess_hunstead2006}.
%% say where error bars come from
%}\label{fig:pic_spec}
%\end{figure*}

The first challenge we encounter in flux calibration is establishing an accurate spectrum of a
calibration reference.  Pictor A is chosen as a calibrator source of necessity;
given the grating lobes of the synthesized beam generated by our
maximum redundancy array configuration, sidelobes of Pictor A and Fornax A dominate 
nearly all beam pointings.  Of these
two sources, Pictor A is the least resolved. 

Unfortunately, the spectrum of Pictor A is particularly poorly characterized in 
the 100--200-MHz band where PAPER operates.  Previous measurements imply that 
the spectrum below 400 MHz deviates substantially
from the power law (with spectral index $\alpha=-0.85$) that holds at higher frequencies \citep{perley_et_al1997}.
There are also serious inconsistencies in the measured flux of Pictor A below 1 GHz
(e.g. \citealt{slee1995}, \citealt{kuhr_et_al1981}, \citealt{large_et_al1981}, \citealt{burgess_hunstead2006}),
and a dearth
of measurements at nearby frequencies in order to establish the spectral slope of Pictor A in
this band.

This shortcoming in the literature has recently been addressed
with observations
from a 64-antenna, single-polarization deployment of the 
PAPER array in a minimum-redundancy configuration more suited for imaging.  The details
of these observations and their calibration are given in \citet{jacobs_et_al2013}, along with
%In order to ensure that errors in
%the beam model did not skew our result, we chose a set of sources near in declination
%to Pictor A that would follow nearly identical tracks through the primary beam.
%%A list of these 
%%sources is given in Table \ref{tab:src_spec}, with the spectra measure by PAPER
%%illustrated in Figure \ref{fig:src_spec}.  These sources were chosen on the basis of brightness and
%%by how well their spectra below 1 GHz exhibit power-law behavior versus frequency.
the resulting spectrum of Pictor A that is shown to be consistent with a single power law
that is best fit by
\begin{equation}
S_\nu= S_{150}\cdot \left(\frac{\nu}{150 {\rm MHz}}\right)^\alpha
\end{equation}
with
$S_{150}=382\ {\rm Jy}\pm5.4$, with $\alpha=-0.76\pm0.01$.
Error ranges indicate 76\% confidence intervals.

%To further improve these estimates, we derive an additional frequency-dependent gain parameter,
%smoothed on 10-MHz scales,
%that brings the spectrum of J2331-416 into closest agreement with a power-law spectrum of
%$S_{150}=31.9\ {\rm Jy}$, $\alpha=-0.69$ that represents the best fit to measurements below 1 GHz
%\citep{slee1995,kuhr_et_al1981,large_et_al1981,burgess_hunstead2006}.
%This additional gain factor was applied to all measured spectra, including that
%of Pictor A.  The expectation is that, since all sources at similar declinations follow similar tracks 
%through the PAPER beam, any errors present in the original measured spectra as a result of differences between
%the simulated and actual beam response should be absent in the ratio between source spectra, and therefore
%should be corrected by the applied gain factor.

%While this 
%disagrees with existing measurements at 160 MHz, our ability to reproduce measurements at 160 MHz
%for dozens of sources other than Pictor A with in-band measurements at 160 MHz, and with
%power-law extrapolations of measurements at nearby frequencies, lends credence to the idea
%that the difference between our measurements and the others at 160 MHz is real.
%Moreover, Stefan et al. (private communication) have performed an independent measurement of the flux
%calibration versus frequency for PAPER using a different set of calibrators,
%and an imaging-based technique. Applying their calibration to Pictor A results
%in a spectrum that agrees with the results presented herein to within 10\%.

\subsubsection{Gain Calibration Augmented by Redundancy}

\begin{figure*}\centering
\includegraphics[width=1.85\columnwidth]{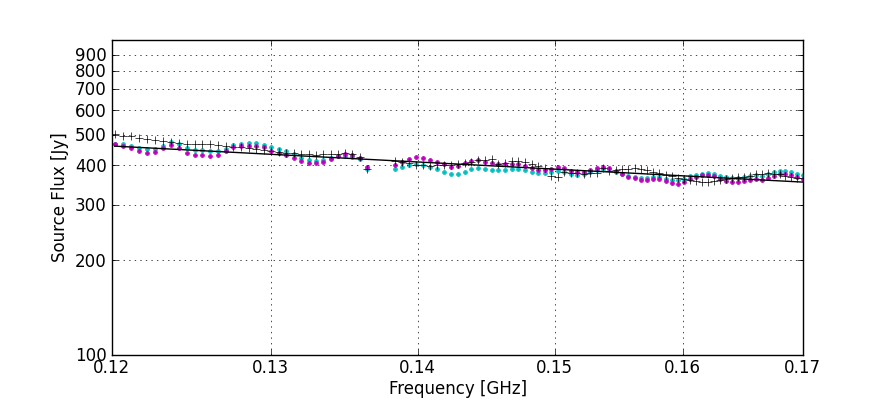}
\caption{Calibrated source spectrum for Pictor A measured using XX (cyan)
and YY (magenta) polarization data from the power-spectrum observations.  Solid lines show power-law 
fits to the data in the range 120--170MHz.
These plots characterize the calibrated flux scale of the measurements used in the power spectrum analysis.
These measurements were calibrated to agree with the Pictor A spectrum in \S\ref{sec:pic_spec} (solid black).
The ripples in the measured spectra are the result of beam sidelobes on Fornax A that are substantially higher for
maximum-redundancy observations than the minimum-redundancy observations (plusses) used in \citet{jacobs_et_al2013}.
}\label{fig:flux_cal}
\end{figure*}

For these maximum-redundancy observations, gain calibration is
derived on the basis of both redundancy and self-calibration.  First,
we correct for the known dependence of amplifier gain and cable attenuation on ambient temperature
using measured temperature data and temperature coefficients that were characterized in
laboratory measurements \citep{parashare_bradley2009} and confirmed in field
measurements \citep{pober_et_al2012}.  
These
corrections correspond to approximately a 5\% variation in amplitudes over the course of the observations.
This is the only time-dependent parameter used in the gain calibration.

Next, redundant self-calibration is used to derive the amplitude of each 
antenna signal path relative to
a single overall bandpass function for each (XX,YY) polarization.  As mentioned in the context
of phase calibration, per-antenna gains
are derived from the ratios of baselines within instantaneously redundant sets, and because the sky signal
cancels in these ratios,
there is no possibility of signal loss associated with redundant gain calibration.

This overall bandpass is then calibrated to the flux scale established in \citet{jacobs_et_al2013}
by phasing to the position of Pictor A
and averaging over all inter-column baselines.
Because Pictor A is resolved at the 20\% level on the longest (300m) East-West PAPER baselines, we 
use per-baseline estimates of resolution weights to down-weight and correct for resolution effects.  These
resolution weights are estimated from the 90cm map of Pictor A presented in \citet{perley_et_al1997}.  Averaged over all
the baselines in the array, the inclusion of resolution effects change the measured flux by less than 5\%.
As in \citet{jacobs_et_al2013}, a fringe-rate filter is applied to the resulting
beam to suppress sidelobes that introduce variations deviating more than 
$\pm$0.1 mHz from the fringe rate of the source in question.

To derive a source spectrum from a drift-scan source profile, we average time samples using
weights derived from a model of
PAPER's primary beam response as an effective approximation of inverse-variance weighting
\citep{pober_et_al2012}, and average over 92 days of drift-scan observations. 
A 9th-order polynomial
is fit over a 100-MHz bandwidth to bring each measured spectrum in accordance with the unpolarized model of the 
Pictor A spectrum described in \S\ref{sec:pic_spec}.
The resulting Pictor A spectrum,
plotted for each polarization in Figure \ref{fig:flux_cal} along with the reference spectrum
derived from minimum-redundancy PAPER observations \citep{jacobs_et_al2013},
characterizes the
flux scale used in the subsequent power-spectrum analysis.

This polynomial bandpass model uses a limited number of terms to prevent the coupling of beam sidelobes, which
are the most likely cause of the residual gain variation in Figure \ref{fig:flux_cal}, into
the overall gain calibration.  While such sidelobes are difficult to distinguish from intrinsic
frequency structure, we err on the side of caution (using a limited number of degrees of freedom)
to avoid introducing any frequency-dependent
gain terms not in the data.  In this cautious approach, it is possible that higher-order frequency
structure in the instrumental bandpass may not be calibrated out.  To address this possibility,
we make the argument that evidence of such structure must be present in the measured
Pictor A spectrum after the application of the bandpass calibration, and to the extent that
some higher-order structure is demonstrably present, this structure can only be ascribed to uncalibrated
structure in the instrumental bandpass if it modulates foreground structure to higher-order modes
in the measured power spectrum.  The
fact that, at the upper limits we report in \S\ref{sec:measured_pspec}, we do not see this structure
argues for its absence.  In any case, this conservative modeling of the PAPER bandpass with as
few degrees of freedom as necessary does not compromise the validity of the upper limits that are reported.

The restricted order of the polynomial used in calibrating the bandpass also serves to
insulate against the signal loss that would result from inadvertently including 21cm EoR
fluctuations in the bandpass model.  We use simulations with many realizations of
low-level white-noise
added to the polynomial bandpass model we fit using Pictor A to quantify the expected
signal loss as a function of $k_\parallel$ in the power spectra measured in \S\ref{sec:measured_pspec}.
In each case, we compare the power spectrum derived from the original injected noise with the
assumed polynomial bandpass model to the 
power spectrum of the injected noise after dividing by a new fit of a 9th-order polynomial to the bandpass.
We find signal loss in the range $-0.06<k<0.06h\ {\rm Mpc}^{-1}$ to be less than 8\%, and signal loss
outside of this range to be below 0.2\% and to decrease rapidly with $|k_\parallel|$.  
This simulation calculates signal loss strictly on the basis of fitting a 9th-order 
polynomial bandpass model to a spectrum containing the 21cm EoR signal, and ignores 
the use of multiple baseline lengths and earth-rotation synthesis in measuring the Pictor A spectrum. 
This averaging of independent modes 
attenuates the coupling of the 21cm EoR signal into the Pictor A spectrum that
is used in the bandpass calibration in proportion to the square root of the number of modes averaged.  Hence, this 
simulation establishes an upper bound on signal
loss that results from bandpass calibration, and shows it to be a small effect for the $k_\parallel$ modes
that we use to set our upper limits in \S\ref{sec:measured_pspec}.  Because polynomial fitting is a linear
process, the fractional signal loss we calculate using a white noise signal model is independent of the
amplitude of the model signal.

% PSA747 used to establish spectrum of Pic A.  Do we report other source spectra here? (no)

Following calibration to a consistent flux scale, XX and YY polarizations for each
baseline are directly summed to form a coarse estimate of the Stokes I parameter. 
This simplistic construction of Stokes I neglects
differences in the beam responses of each linear polarization, and as a result,
contains contributions from Stokes Q away from beam center \citep{jelic_et_al2010,moore_et_al2013}.  Nonetheless,
for analysis of a single baseline type, correcting for direction-dependent gains is highly non-trivial
at best, and we attempt no such correction here.
On the basis of the asymmetry of the PAPER beam model
under 90$^\circ$ rotation, we estimate that this construction of Stokes I will contain approximately
4\% of Stokes Q as well \citep{moore_et_al2013}.  For a more thorough treatment of polarization 
in the context of drift-scanning
arrays with delay-spectrum analysis, we refer the reader to \citet{moore_et_al2013}.
% this section may need more nuanced discussion.  Add Stokes Q? (no)

%lstbin.py -a cross -C psa898_v002 --lst_res=42.95 --lst_rng=$LST_1HR --tfile=600
%xrfi_simple.py --dt=3 -c 74,152,153,168 
%xtalk3.py 
%apply_cal.py -C psa898_v003 

\subsection{Wideband Delay-Spectrum Foreground Separation}
\label{sec:wideband_dspec}

%pspec_prep.py -C psa898_v003 -a cross --window=blackman-harris --nogain --nophs --clean=1e-9 --horizon=15 

The next major step in our data reduction pipeline is to remove a best estimate of 
foreground emission arising from the galactic synchrotron and extragalactic point sources.
This is an essential step, as foreground emission exceeds the expected level of EoR emission by
9 orders of magnitude \citep{pober_et_al2013}, and can conceal low-level RFI and crosstalk systematics.  Moreover,
as described in P12b, smooth-spectrum foreground
emission can corrupt higher-order $k$-modes in power-spectrum measurements as a result
of sidelobes that arise from RFI flagging and the finite (windowed) bandwidth used in the line-of-sight
Fourier transform.  Although power-spectrum measurements that aim to constrain 21cm reionization
must be limited to $\sim\!$10 MHz to avoid significant signal evolution within the band,
foreground emission is generally coherent over the entire 100-MHz band of PAPER observations.  

\begin{figure*}\centering
\includegraphics[width=1.85\columnwidth]{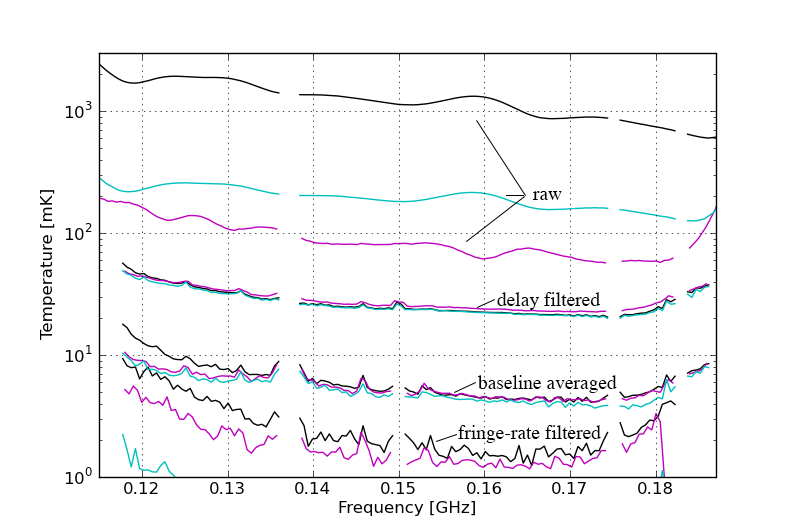}
\caption{
RMS brightness temperatures of raw 
(black), frequency-differenced (magenta), and the
time-differenced (cyan) visibilities in various stages of analysis.
The upper triplet of curves correspond to raw visibilities for individual 30-m east-west baselines, 
averaged coherently in 43-second LST bins over 20 days with a channel bandwidth of 490 kHz, and
averaged incoherently over the set of 28 identical baselines.
The second triplet of curves from the top
corresponds to visibilities 
after the application of the wide-band delay filter described in \S\ref{sec:wideband_dspec}.
The third triplet shows the result of averaging delay-filtered visibilities coherently over the set
of 28 baselines.
The lowest triplet shows
the same baseline-averaged visibilities after the application of 
the fringe-rate filter described in \S\ref{sec:fringe_rate_filtering}.
The lower cyan curve is significantly lower than the others as a result of the near-orthogonality, in
fringe-rate domain,
of a high-pass time-differencing filter and the low-pass fringe-rate filter; it is not a valid
representation of the noise.  Scaling the wide-band delay filtered curves by the effective integration
time and bandwidth, we determine a system temperature of 560 K at 160 MHz.
}\label{fig:noise_vs_fq}
\end{figure*}

A delay-domain filter applied over a wide bandwidth can achieve a much sharper degree
of foreground isolation than is possible in any sub-band, and this can be used to effectively
model and remove a smooth foreground model from each sub-band of interest.  Following the procedure
outlined in P12b and implemented in \citet{pober_et_al2013}, we perform a delay transform of each integration over
the entire observing band, according to the equation
\begin{equation}
\tilde{V}(\tau)=\int{W(\nu)\cdot S(\nu)\cdot V(\nu)e^{-2\pi i\nu\tau}~d\nu},
\label{eq:dtransform}
\end{equation}
where $V(\nu)$ is the measured visibility as a function of spectral frequency $\nu$, $\tau$ is delay --- the 
Fourier complement to $\nu$, $W(\nu)$ is a Blackman-Harris windowing function \citep{harris1978}
used for minimizing sidelobes and band-edge effects, $S(\nu)$ encodes the frequency-dependent sample weights 
that result from RFI flagging, and `$\sim$' denotes the delay transformation.  Following the procedure outlined in \citet{parsons_backer2009}, we deconvolve
the delay-domain kernel that results from the product $W(\nu)\cdot S(\nu)$ using a CLEAN-like iterative forward modeling deconvolution algorithm
\citep{hogbom1974} in delay domain,
restricting model components to fall inside of 15 ns beyond the horizon limit of each baseline.
We chose a 15 ns standoff from the horizon to suppress the first level of supra-horizon emission that
results from the inherent width of $\tilde{S}(\tau)$.
Because the sampling function is known exactly, the restriction of model components to low-order delay modes
ensures that the deconvolution process cannot introduce power at higher-order delay modes beyond what was
scattered there by sidelobes of the sampling function.  Moreover, 
because of the least-square metric used in evaluating CLEAN models for divergence, any
delay-domain residuals associated with the sampling function will be unbiased.

The resultant CLEAN model is subtracted from $\tilde{V}(\tau)$ to effect a delay-domain filter
suppressing smooth-spectrum emission.
On the 30-m, nearly east-west baselines used in this
analysis, in a band centered at 164 MHz, this corresponds to a cutoff at $k=\pm0.06\ h\ {\rm Mpc}^{-1}$.
This delay filter is solely responsible for
-60 dB of suppression (in mK$^2$ power) of the foreground signal seen between the unfiltered
signal (top black curve) and the final residual (lowest black curve) shown in Figure \ref{fig:noise_vs_fq}.
More importantly, in the narrow-band delay transformation that forms the basis of power spectrum measurements
in \S\ref{sec:dspec_crossmult}, this filter heavily suppresses the coupling of delay modes beyond the horizon to 
the foreground-dominated modes within the horizon.  There is a small amount of signal loss associated
with this filtering. Using the simulation methodology described in \S\ref{sec:cov_diag},
we compute 
this signal loss to be 4.8\% for the two modes on either side of the horizon limit, 1.3\% for the next modes
outside of that, and  $<$0.015\% elsewhere.\footnote{That the signal loss is relatively small outside the horizon is perhaps unsurprising.  To achieve a
coherent reduction of the cosmological signal outside the horizon, it is necessary for delay-space sidelobes
to couple signal from outside the horizon to modes within the horizon, which are then forward modeled to
give outside-the-horizon sidelobes to subtract from the original signal.  This sort of coherent reduction represents two factors of sidelobe attenuation on top of an already small cosmological signal, and is therefore negligible.}
  The cost of this signal loss is far outweighed by
the benefits of suppressing couplings to foreground-dominated modes where, despite the low coupling strength,
the sheer magnitude of the foreground signal creates a dominant bias.

%xrfi_simple.py -n 3 -c 0_35,176_202 
%xtalk3.py 
%sum polarizations and baselines
%xrfi_simple.py -n 3 -c 101,102

After the removal of the bulk of the foregrounds, the signal remaining for each baseline is visibly
noise-dominated. 
%noise-dominated.  We sum the dynamic spectra of the 28 redundant baselines in columns of four, 
%leaving seven independent time-series of frequency-dependent visibilities, each with quadruple
%the sensitivity of a single baseline.
At this point, we apply a final pass of RFI excision, flagging 3$\sigma$
outliers, and proceed to crosstalk removal to remove systematics that were previously undetected
beneath the bright, smooth-spectrum emission.  Crosstalk removal proceeds by subtracting the
1-hr time-average of the visibilities for each baseline from each integration.  This process,
described in \citet{parsons_et_al2010}, distinguishes oscillating fringes associated with 
celestial emission from the static phase bias associated with crosstalk. This crosstalk-removal
filter essentially constitutes a high-pass fringe-rate filter, as described in \S\ref{sec:fringe_rate_filtering}.
The width of the stop-band of the crosstalk filter is much narrower than low-pass fringe-rate filters described in
that section.

Nighttime data are then averaged in LST over the 92-day observation using 43-second
time bins that match the integration interval of the data after the compression step described
in \S\ref{sec:preprocessing}.
In the LST binning process, sporadic non-Gaussian events that evade the RFI flagging process
can easily skew the average value in an LST bin away from the median value.  To avoid bias from
non-Gaussian tails, we flag the 10\% of data in each LST bin, for each frequency channel, that deviate
most from the median value, and exclude them from the computed average for that LST bin.  This last
flagging step suppresses non-Gaussian outlying events without biasing the computed average in an LST bin.
We simulate this median filtering process for samples in a normal distribution around an underlying random signal
assigned to each LST bin, and verify that the computed variance of the underlying signal is not biased by
the median filter, regardless of the level of the signal with respect to the noise.
We also use this simulation to verify that bootstrapping correctly reproduces the range of variation
in the final estimator as expected for Gaussian statistics, and that the bootstrap error bars are unaffected
by the median filter.  These simulations indicate that the LST binning process described above
has no associated bias or signal loss for recovering the cosmological EoR signal.

\subsection{Fringe-Rate Filtering}
\label{sec:fringe_rate_filtering}
%fringe_rate_filter.py -C psa898_v002 -a cross --clean=1e-2 --minfr=-1 --fr_frac=1.2 

In the last step prior to forming power spectra,
we apply a fringe-rate filter to effect time-domain integration,
using the effective time interval that a baseline measures a single $k$-mode to integrate coherently
(with noise decreasing
as $\sqrt{t}$, in units of mK), before measurements at different times represent independent modes
that must be squared before further integration (with noise now decreasing as $\sqrt{t}$, in units of mK$^2$).

One way of handling this additional integration is via gridding in the $uv$-plane.
Each measured visibility in a wide-field interferometer represents the integral over a kernel in
the $uv$-plane that reflects the primary beam of the elements \citep{bhatnagar_et_al2008,morales_matejek2009} and the $w$ component 
of the baseline \citep{cornwell_et_al2003}.  As noted in
\citet{sullivan_et_al2012} and \citet{morales_matejek2009},
in order to optimally account for the mode-mixing introduced by these kernels, gridding kernels must be
used that correctly distribute each measurement among the sampled $uv$-modes, such that, in the ensemble average
over many measurements by many baselines, each $uv$-mode becomes an optimally weighted estimator of the actual
value given the set of measurements.

However, this approach has a major shortcoming when applied to maximum-redundancy array configurations.
In order
to maximize sensitivity, such
configurations are set up to deliberately sample identical visibilities that reflect the same 
combinations of modes in the $uv$ plane, with few nearby measurements (P12a).  As a result,
such array configurations tend to lack enough measurements of different combinations of $uv$ modes
to permit the ensemble average to converge on the true value.  Said differently, maximum-redundancy
array configurations tend to produce measurement sets that, when expressed as linear combinations
of $uv$-modes of interest, are singular.

Our alternative approach avoids this and many of the difficulties outlined
in \citet{hazelton_et_al2013} by applying a carefully tailored
fringe-rate filter to each time series of visibility spectra.  As outlined in Appendix \ref{app:data_compression}
in the context of data compression, we
take the Fourier transform of the time series in each channel and apply a low-pass filter that preserves
fringe-rates that geometrically correspond to sources rotating on the celestial sphere.  
For a planar array with transit observations, fringe-rates vary according to declination, with fringe rates
reaching a maximum ($f_{\rm max}$) at 
$\delta=0^\circ$, decreasing to 0 at $\delta=-90^\circ$, and for an array such as PAPER deployed near
-30$^\circ$ S latitude, reaching a minimum of $\approx-f_{\rm max}/2$ at $\delta=-60^\circ$ on
the far side of the south celestial pole.
In order to
avoid introducing undesirable frequency structure, we apply the same filter, tuned to the width
set by the highest frequency of the sub-band used in the
power spectrum analysis described
in \S\ref{sec:dspec_crossmult}, to each channel,
even though maximum fringe-rates are generally frequency-dependent.
%Since fringe rates are a
%natural basis for celestial emission
%over short time intervals, fringe-rate
%filters can be narrowly tailored to the geometric bounds of a baseline.
%In contrast, simply summing integrations over an equivalent time interval 
%(corresponding to a sinc filter in fringe-rate space) will tend to significantly suppress emission at
%fringe rates that geometrically correspond to the sky.
In a future paper, we will explore the idea
of employing fringe-rate filters that purposely down-weight fringe-rate modes on the sky according to
the expected signal-to-noise ratio in each mode.  Such filters would essentially correspond to a
one-dimensional implementation of the inverse primary beam $uv$-gridding discussed in \citet{morales_matejek2009},
and have many features in common with m-mode synthesis described in \citet{shaw_et_al2013}.

Since thermal noise scatters equally into all fringe rate bins, applying a filter
that passes only fringe rates corresponding the celestial emission has the effect of de-noising the data.
We apply such a filter to the data, choosing the bounds of the filter to match the geometric
bounds set by a 30-m east-west baseline, according to the equation
\begin{equation}
f_{\rm max} = \frac{|\b_{\rm eq}|}{c} \omega_\oplus \nu,
\label{eq:fringe_rate}
\end{equation}
where $f_{\rm max}$ is the maximum fringe rate, 
$\b_{\rm eq}$ is the baseline vector projected parallel to the equatorial 
plane, $c$ is the speed of light,
$\omega_\oplus$ is the angular frequency of the Earth's rotation,
and $\nu$ is the spectral frequency. 
% 780 = 30e2 / c * 2pi / 86164 * .174, where .174 is chosen because want to not attenuate sky over entire band
At 174 MHz (the highest frequency in a 20-MHz window centered on 164 MHz that is used in \S\ref{sec:dspec_crossmult}),
$f_{\rm max}=1.27$ mHz, corresponding to a fringe period of 788 s.  Hence, the fringe-rate filter that is
applied passes fringe-rates in the range $-0.63<f<1.27$ mHz.  The width of this filter corresponds in 
% reviewer wants clarification on exact FR limits that give 525 as opposed to 500s.
sensitivity to an effective integration time of 525 s.
We note that
this filtering could have been applied during the data compression described in \S\ref{sec:preprocessing},
but was implemented separately to enable the compression to work uniformly
over all baselines in the array without additional information about antenna location.

After applying this filter,
we transform the data back to time domain in preparation for forming power spectra via the delay transform.
It should be noted that, in time domain, the data are now heavily over-sampled; adjacent samples are no longer
statistically independent.  Hence, when averaging power-spectra versus time,
noise will not beat down according to the strict number in samples, but rather, according to
the actual number of statistically independent samples underlying the time series.

\subsection{Narrow-band Delay-Transformation, and Cross-Multiplication}
\label{sec:dspec_crossmult}
% Cross-multiplication of baselines
%pspec_redmult_cov.py -a 0_16,8_16,8_24,4_24,4_20,12_20,12_28 -c 110_149

In the final stage of our core analysis, 
we
apply the delay transform in Equation \ref{eq:dtransform} to a 20-MHz band centered at 164 MHz, again
using a 
Blackmann-Harris window that yields an effective (noise-equivalent) bandwidth of 10 MHz,
centered at redshift $z=7.7$.  
In this paper, we limit our analysis to a single cosmological bin where, out of the entire
PAPER observing bandwidth, the noise 
and foreground residual in Figure \ref{fig:noise_vs_fq} is lowest.
As discussed in P12b, the $\tau$-modes that result from the delay transform
have a peaked response in $k_\parallel$ that allows each $(u,v,\tau)$-mode to be interpreted 
as sampling the mode $\k=2\pi(u/X,v/X,\tau/Y)$, with $X$ and $Y$ representing conversion factors 
from angle and frequency to comoving distance, respectively.  As such, we use $k_\parallel$ and
$\tau$-modes interchangeably in the following discussion.

In contrast to the wide-band delay filtering described in \S\ref{sec:wideband_dspec}, where a
CLEAN-like deconvolution with a restricted range of model components was used to suppress the 
sidelobes of smooth-spectrum emission
resulting from unsmooth sampling functions, we do not deconvolve covariances in
$\tilde{V}(\tau)$ introduced by windowing and sampling functions. 
Firstly, in the band chosen
for this analysis, the sampling function $S(\nu)$ in equation \ref{eq:dtransform} is nearly unity at
all frequencies, thanks to the exquisite RFI environment at the Karoo site in South Africa.
Secondly, the covariances between nearby $\tau$ modes introduced by the windowing function go hand-in-hand with
the suppression of covariances between modes at larger separations.  Hence, we defer the removal of
covariances introduced by the windowing and sampling functions at this stage until the covariance diagonalization
process described in \S\ref{sec:cov_diag}.

From these delay-transformed visibilities,
we construct unbiased estimators of the power spectrum, $\widehat P(\k)$,
by cross-multiplying delay spectra between 
the seven baseline groups constructed in \S\ref{sec:wideband_dspec}, 
and summing over all cross-multiples, according to the equation:
\begin{equation}
\widehat P(\k_{t\tau}) = \left(\frac{\lambda^2}{2k_{\rm B}}\right)^2\frac{X^2Y}{\Omega B}
\left\langle{\tilde V_i(\tau,t) \tilde V_j^*(\tau,t)}\right\rangle_{i<j},
\label{eq:pspec_cosmo}
\end{equation}
which follows from equation 12 of P12a, with $\lambda$ being the observing
wavelength, $k_{\rm B}$ is Boltzmann's constant, $X^2Y$ is a cosmological scalar with units
of $\frac{h^{-3}\ {\rm Mpc}^3}{{\rm sr}\cdot {\rm Hz}}$, $\Omega$ is the angular 
area\footnote{
As described in detail in Appendix \ref{app:beam_area}, the angular area used to normalize
high-redshift 21cm power spectrum measurements (e.g., $\Omega$ in Equation \ref{eq:pspec_cosmo}) is proportional
to the integral of the squared beam power over angular area ($\Omega_{\rm PP}$; equation \ref{eq:beam_squared}).
This contrasts the standard beam area ($\Omega_{\rm P}$; equation \ref{eq:beam_area}) that is
used to relate flux density to a brightness temperature.
Since Equation \ref{eq:pspec_cosmo} 
relates a measured visibility in units of brightness
temperature to $P(\k)$, a factor of $\Omega_{\rm P}^2$ has already been
applied to convert Jy to mK.  In this case,
$\Omega$ indicates the remaining factor of $\Omega_{\rm PP}$, which for PAPER is 0.31 sr.},
% mention  that windowed bandwidth also must be calculated? Actually, this is just a detail of how ifft, works.
$B$ is the bandwidth, $\langle\dots\rangle_{i<j}$ indicates the ensemble average
over instantaneously redundant baseline measurements indexed by $i,j$,
and $\tilde V(\tau,t)$ is the delay-transformed visibility,
expressed in terms of delay $\tau$ and time $t$.
We use $t$ as a subscript on $\k$
to denote the different modes sampled by a baseline as the sky rotates, and $\tau$ to indicate
the dependence of $\k$ on the delay mode in question.

Equation \ref{eq:pspec_cosmo} represents the diagonal-covariance limit of an optimal quadratic estimator
\citep{dillon_et_al2013a,liu_tegmark2011}, which we review in detail and adapt to
delay-transformed visibilities in Appendix \ref{app:cov_diag}.  The diagonal limit generates
optimal power spectrum estimators in the simple case
of ideal signal covariance among statistically independent $\tau$-modes.
In theory, the translation-invariance of the cosmological signal (along with
the interchangeability of $\tau$ and $k$) ensures that the signal covariance is diagonal.
Over narrow frequency bands such as the one used for these measurements, a similar invariance
(and thus, diagonality) along the frequency direction
should hold for the noise and foregrounds \citep{dillon_et_al2013a}.
In practice, we encounter substantial systematics whose covariance
deviates from this ideal.  In \S\ref{sec:cov_diag}, we examine how off-diagonal covariances
arising from systematics can be suppressed in a straightforward way and incorporated
into Equation \ref{eq:pspec_cosmo} to produce power spectrum estimators that are
to leading order, optimal.  Because this later analysis follows the same trajectory as
the analysis presented, we will first bring the current analysis to completion before
examining how it is modified to incorporate off-diagonal covariances.

For $N$ redundant samples of a $\k$-mode, $N(N-1)/2$ cross-multiples are constructed.  At this
stage, we do not use $i>j$ pairings, choosing instead to preserve the imaginary component of the
power spectrum estimator, even though $P_{21}(\k)$ is real-valued, as a diagnostic.
In the final power spectrum estimate, we drop the imaginary component, which is 
equivalent to including the $i>j$ baseline pairings.  Hence, we ultimately have $N^2-N$ measurements of
$P(\k)$, with noise decreasing proportional to $N$ in mK$^2$ units, as would be expected for coherently
integrating redundant samples before squaring them.

Next, we average over measurements of independent $\k$-modes that statistically reflect the same
underlying power spectrum, to produce our best power-spectrum estimate, 
\begin{equation}
\widehat \Delta^2_{21}(k) = \frac{k^3}{2\pi^2}\left\langle \widehat P(\k_{t\tau})\right\rangle_{|\k_{t\tau}|=k},
\end{equation}
where the three-dimensional symmetry of the power spectrum is invoked to average over
all independent measurements of modes in a shell of $|\k|=k$, with independent measurements
indexed here by $t$.  As described in \S\ref{sec:fringe_rate_filtering}, the number of independent modes
that are averaged (with noise decreasing with number of modes, $M$, as $\sqrt{M}$ in mK$^2$ units; see P12a) is 
determined
by overall observing window and the number of fringe-rate bins that are 
preserved in the fringe-rate filtering process.
Noise levels are estimated using bootstrapping, as described in \S\ref{sec:bootstrap}.
Since we have not decimated the number of integrations to the critical sampling rate corresponding to the 
width of the applied fringe-rate filter, $M$ is {\it not} the number of integrations.  However,
we are free to average the power spectrum estimates for each integration, even though nearby samples
do not have statistically independent noise, understanding that noise will decrease according to the number
of underlying independent samples.

\begin{figure*}\centering
\includegraphics[width=1.85\columnwidth]{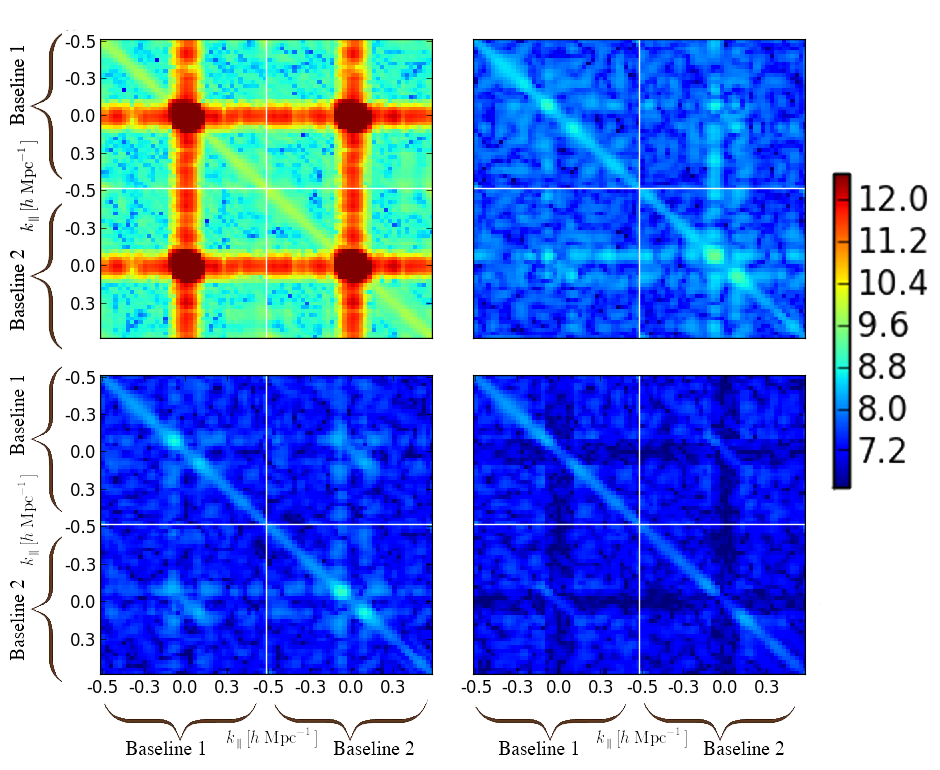}
\caption{
Upper-left: The covariance matrix of delay modes from two representative baselines generated by applying
a narrow-band delay transform (\S\ref{sec:dspec_crossmult}) to data without the application of a wide-band delay filter;
upper-right: Similar to upper-left, but applied to data after the application of a wide-band delay filter
(\S\ref{sec:wideband_dspec});
lower-left: Similar to upper-right, but after removing per-baseline off-diagonal covariance terms that deviate
from the average (\S\ref{sec:cov_diag});
lower-right: Similar to lower-left, but after removing off-diagonal covariance terms common to all baselines
associated with the central seven delay modes.  This final step attenuates the expected level of the
reionization signal in these central modes by $\sim$30\%, but only attenuates other modes by $\sim5$\%.
}\label{fig:cov}
\end{figure*}

We also apply this same process for constructing power-spectrum estimators
to data that include smooth-spectrum foreground emission before it was suppressed
with the wide-band delay filter described in \S\ref{sec:wideband_dspec}.  
While these data are not expected to yield accurate measurements of $P(\k)$ at higher $k$
owing to sidelobes of data flagging and low-level systematics, as mentioned previously, they do yield
effective measurements of the bright foreground emission that falls within the limits of the wide-band
delay filter (top-right panel, Figure \ref{fig:cov}).  At the baseline length used in
this analysis, this emission is dominated by bright
point sources such as Pictor A and Fornax A \citep{parsons_et_al2010,jacobs_et_al2011}.
In order to present a complete picture
of the power spectrum of emission on the sky
 that is valid at all $k$-modes, 
we stitch together the final power spectrum using measurements
derived from foreground data at $k$-modes where emission is suppressed by the wide-band
delay filter, and using measurements from the foreground-suppressed data elsewhere.

\begin{figure*}\centering
\includegraphics[width=1.85\columnwidth]{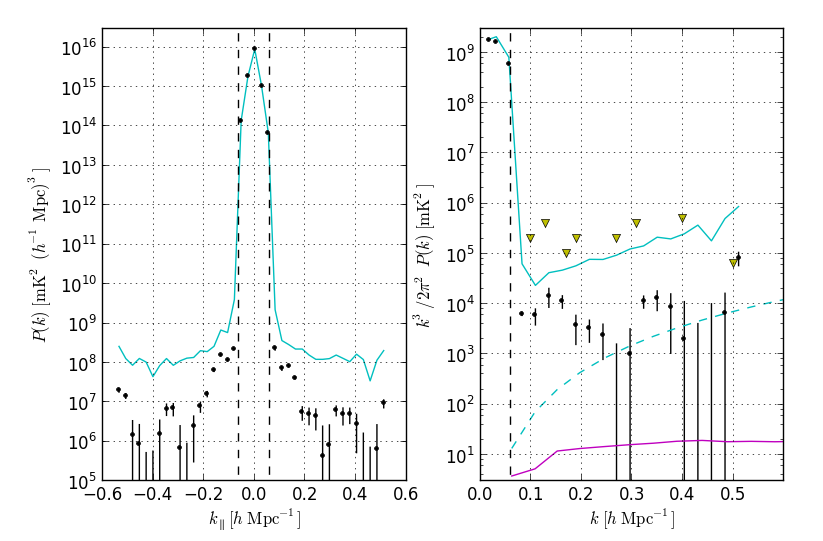}
\caption{
Power spectra at $z=7.7$ derived from the 92-day PAPER observation described in \S\ref{sec:observations}.
In both panels, solid cyan depicts 2$\sigma$ upper limits derived from PAPER observations without the 
removal of off-diagonal covariance terms, and black indicates the final measured power spectrum with 2$\sigma$ confidence 
intervals.
Because of windowing in the delay transform (\S\ref{sec:dspec_crossmult}), adjacent measurements
are $\sim$50\% correlated.
The measurements in the right panel are weighted averages of positive/negative $k_\parallel$ contributions.
The horizon limit (vertical dashed) illustrates the boundary within 
which emission has been filtered out in delay space and re-inserted after the formation of power spectra.
Dashed cyan illustrates the predicted noise power spectrum from \citet{parsons_et_al2012a} for a system
temperature of 560 K.
The yellow triangles indicate 2$\sigma$ upper limits reported in \citet{paciga_et_al2013} at $z=8.6$.
Magenta illustrates a fiducial model at 50\% ionization \citep{lidz_et_al2008}.
%and using $k_\perp\approx0.01h\ {\rm Mpc}^{-1}$ for the 16-wavelength baselines used in this
%analysis. 
At $k\approx0.27\ h\ {\rm Mpc}^{-1}$,
we report
an upper limit on $\Delta_{21}^2(k)$ of 1660 mK$^2$ with 2$\sigma$ confidence.
%The exact
%cause of the asymmetry between positive and negative $k_\parallel$ is still under investigation;
%such an asymmetry can arise from bright foregrounds being closer in delay space to the eastern
%horizon, or from polarization leakage from signals with positive rotation measures.
}\label{fig:pk_k3pk}
\end{figure*}

The result of the analysis up to this point is the power spectrum illustrated 
by the cyan curve in Figure \ref{fig:pk_k3pk},
which exhibits residual off-diagonal covariance structure that is illustrated in the top-right panel
of Figure \ref{fig:cov}.

\subsection{Suppression of Off-Diagonal Covariances Arising from Systematics}
\label{sec:cov_diag}

The off-diagonal covariances shown in the upper-right panel of
Figure \ref{fig:cov}, which are estimated empirically from measured data,
are indicative of systematics corrupting what should
nominally be nearly statistically independent estimators of $k$-modes of reionization.  This diagnosis is 
reinforced 
by the substantial variation of the covariance structure between different baselines;
we expect the sky signal, whether from foregrounds or reionization, to be common to all baseline
cross-products (as in the top-left panel of Figure \ref{fig:cov}).  Hence, we now turn to investigating
techniques for removing these systematics while retaining, to the greatest extent possible, the 
reionization signal we are after.  As we describe briefly here and in detail in Appendix \ref{app:cov_diag},
this can be achieved by using the quadratic estimator formalism \citep{dillon_et_al2013a,liu_tegmark2011}.

As derived in Appendix \ref{app:cov_diag}, the special case of
applying optimal quadratic estimators to delay-domain visibilities can, to leading order,
 be approximated as a modification of the visibility.  
In brief, knowledge of the covariances allows off-diagonal leakages between delay bins to
be predicted and removed.
If we take the output of the
off-diagonal covariance removal process described in Equation \ref{eq:cov_diag} 
a perturbed visibility, $\tilde{V}^\prime(\tau)$ from Equation \ref{eq:vtildeprime}, such that 
%\begin{equation}
%\widehat{p}_\alpha\equiv\left(\frac{\lambda^2}{2k_{\rm B}}\right)
%\sqrt{\frac{X^2Y}{\Omega B}}\tilde{V}^\prime(\tau_\alpha),
%\end{equation}
then for this modified visibility, Equation \ref{eq:pspec_cosmo} encodes the remainder of the optimal
estimator formalism.

There are, however two subtle issues to address in applying the off-diagonal covariance
removal process described in Appendix \ref{app:cov_diag}.  The first problem pertains to our limited ability to empirically deduce the 
true covariances present in our data.  Because we have a limited number of independent samples
from which to estimate covariances in the data, we measure residual
noise and signal covariances that, in an infinite sample set, would not exist.  If we naively
subtract these residual covariances, we over-fit the noise and suppress the 21cm signal.

There is a simple prescription for avoiding this problem, given that, as the top-right panel
of Figure \ref{fig:cov} illustrates, the dominant systematics we see in the covariance of
$k$-modes exhibit a baseline-dependent structure.  Since celestial emission, whether from 
foregrounds (top-left panel, Figure \ref{fig:cov}) or 21cm reionization, should
have identical structure among redundant baselines, we may subtract the average covariance
among many baseline cross-products (the ``panels'' in each subplot of Figure \ref{fig:cov}) from each individual
cross-product to remove the residual signal covariance and leave behind only the covariance
of noise and systematics from which off-diagonal covariances can be removed --- even
if they are residuals that would not be present in the true covariance matrix --- without
attenuating the desired celestial signal.  This lossless removal of systematics is
another substantial benefit of the highly redundant configurations presented in P12a.

We apply this technique for each baseline cross-product, using all baselines
except the two being cross-multiplied for estimating the average covariance in
order to avoid, to the greatest extent possible, coupling baseline-dependent
systematics into each estimate of the common-mode covariance.  Iterating
this process a modest number of times (two or three), produces an improvement
in the results as initial baseline-specific systematics are first removed from
the baseline, allowing an improved estimate of the average covariance to be
subtracted, such that the remainder of the baseline-specific systematics are
removed.  Further iteration is not necessary, as the process rapidly converges
to the result shown in the bottom-left panel of Figure \ref{fig:cov}.

This process, though quite effective, introduces a second issue
that must be addressed.  In Equation \ref{eq:pspec_cosmo}, we were careful to exclude
products between the same baseline in order to avoid incurring the noise bias that
would result.  The benefits of avoiding this bias far outweigh the slight improvement
in sensitivity that including such ``auto-products'' would produce.  Unfortunately,
by using the average of many baselines to estimate and subtract an average covariance
from each baseline cross-product, and then subtracting off-diagonal terms in the residual,
we couple the baseline-averaged noise into the data
from each baseline.  The result is a low-level residual noise bias approximately equal to
the noise in the power-spectrum estimate.

The most straightforward approach we have found for eliminating this residual noise bias,
other than direct subtraction (which introduces additional complications), is to 
%divide baselines into two separate groups of approximately equal size.
divide baselines into four separate groups of approximately equal size.
Within each group, we apply the off-diagonal covariance removal process, including the subtraction
of an average covariance, using data only from baselines within that group.  Then, to avoid
%incurring a noise bias, we use only cross-products of baselines between these two groups to
incurring a noise bias, we use only cross-products of baselines between these four groups to
estimate $P(\k)$.  By excluding intra-group cross-products, this approach sacrifices
%a factor of approximately $\sqrt{2}$ in sensitivity (in mK$^2$), but as before, we find the benefits
a factor of approximately 15\% in sensitivity (in mK$^2$), but as before, we find the benefits
of avoiding noise bias to outweigh the loss in sensitivity.

The last step in the process of suppressing off-diagonal covariances in our data is,
for select modes, to relax the constraint of not subtracting a
baseline-averaged covariance for selected modes, even if it results in overfitting the noise 
and the signal suppression
that is associated with it.  Particularly, we note the interior seven $k$-modes
(the five inside of the horizon limit shown in Figure \ref{fig:pk_k3pk}, as well as the
first modes beyond this limit on either side) that we measure are more than an order of magnitude
brighter than other modes, and are so corrupted by smooth-spectrum foreground emission that, barring
a heroic effort aimed at modeling and removing these foregrounds, they are unlikely to be useful
for constraining high-redshift 21cm emission.  We find it advantageous to remove all
off-diagonal covariances associated with these modes, even if it means overfitting the noise.
The result of this process is shown in the bottom-right panel of Figure \ref{fig:cov}.

Since we have overfit the noise in this final step, it now becomes necessary to investigate how
we have affected the 21cm signal that we aim to measure, since failure to account for
signal attenuation can lead to erroneous constraints.
We use Monte Carlo simulations (see, e.g., 
\citealt{masui_et_al2013}, \citealt{paciga_et_al2013}, and \citealt{switzer_et_al2013}) to estimate
the expected signal attenuation through the analysis chain from \S\ref{sec:dspec_crossmult} onward,
as illustrated by the red data-flow path in Figure \ref{fig:data_flow}.
Because our analysis does not model out modes of arbitrary shape (such as occurs in
principal component analysis), and because delay filtering and covariance removal operates in the same
space in which the power spectrum is measured, the simulations needed to characterize signal loss
for our analysis are relatively straightforward.
In our simulations, we apply the analysis pipeline used on the data, including 
off-diagonal covariance suppression, to battery of simulated datasets that contain both a randomly generated
sky signal and thermal noise.  In each dataset, the model sky 
signal consists 
of a 
spatially
random, flat-spectrum signal common to all baselines that rotates with the celestial sphere, modeling
the
output of the fringe-rate filtering step that precedes the narrow-band delay transform in
Figure \ref{fig:data_flow}. The noise that is added to this sky model is a Gaussian random signal
unique to each baseline, to which a fringe-rate filter is applied.\footnote{
Because the final result of this paper is a set of upper limits and not a detection of the cosmological
power spectrum, it is not necessary to include a cosmological signal at the theoretically expected level
in our simulations.  The simulations only need to quantify signal loss at the current amplitude levels, such
that if there were a signal entering at the level of our upper limits, we would have robustly detected it.  Since
the upper limits described in \S\ref{sec:measured_pspec} are dominated by residual foregrounds
and thermal noise, it is unnecessary to include the completely subdominant theoretical signal.  We
inject signal amplitudes for $P(k)$ in the range $10^6$--$10^7$ mK$^2$, which may be compared with
the upper limits shown in the left panel of Figure \ref{fig:pk_k3pk}.}

We examine two cases in our simulations.  In the first, we
apply the off-diagonal covariance removal process using empirically determined
covariances internal to the simulated data.
This case effectively treats the simulated signal as if it were the data,
and investigates the effects of overfitting noise in the data
used to produce the covariance matrices.
In the second case, we apply the exact diagonalization matrices used to remove covariance terms in the measured
data to the simulated signal.  This case examines the effect that the operations
applied to the data would have on a simulated signal that is uncorrelated with the data, as would be the case
for a faint signal that is subdominant to foregrounds and instrumental systematics in the data.
These two cases serve to characterize any 
combination of signals that are correlated and/or uncorrelated with the data, and hence accurately bound
the signal loss to which a potential 21cm reionization signal would be subject in this analysis.

In both simulation cases, the amplitude of each $k$-mode in the output 
power spectrum of the recovered signal is compared to the known input amplitude of the simulated signal, averaged
over 1000 independent simulations.
In the first case, results indicate that
removing the off-diagonal covariances associated with the central seven modes results in a $\sim5\%$
reduction in signal amplitude (in mK$^2$) for other modes, and a $\sim30\%$ reduction for the seven
modes in question.  We compensate by dividing by these signal attenuations when reporting the power
spectrum limits in Figure \ref{fig:pk_k3pk}.  In the second case, which tests the effects of our 
exact analysis on the noise statistics of an uncorrelated signal,
we find that noise that is uncorrelated with the data is
not significantly affected by this analysis.

We note that none of the analysis described above constitutes formal model subtraction.
Because we estimate covariances from data, it is possible to over-fit noise and subtract signal,
but we take steps to avoid this, and to compensate for signal attenuation where subtraction
does happen by estimating signal attenuation via Monte Carlo simulations.  
If the full covariances were known a priori, this analysis would be strictly lossless 
\citep{tegmark1997}, and as shown in Appendix \ref{app:cov_diag}, optimal to leading order.

\subsection{Estimating Residual Noise}
\label{sec:bootstrap}

One consequence of the process for removing off-diagonal covariances described above is
that baseline-specific covariances, including residual covariances from thermal noise, 
are heavily suppressed.  As a result, it is not possible to estimate noise residuals
simply from the variation between baseline cross-products at the conclusion of this process.
Instead, we use bootstrapping of baseline samples at the input of the off-diagonal
covariance removal process to estimate error in the output power-spectrum estimate.

There are, however, a few subtleties in applying bootstrap resampling in this context.
One complication is that, as described in \S\ref{sec:cov_diag}, in order to avoid incurring
a noise bias, it is vital that the removal of off-diagonal covariances proceed on independent
groups of baselines.  To reflect that fact that our power spectrum estimator is
unbiased, it is important to restrict the resampling with replacement
%to avoid including data from the same baseline into the two independent baseline
%groups.  This is done by first randomly assigning baselines uniquely to one of the two groups, and then
to avoid including data from the same baseline into the four independent baseline
groups.  This is done by first randomly assigning baselines uniquely to one of the four groups, and then
applying sampling with replacement only within each of these groups.  At each iteration of resampling,
the assignment of baselines to groups is randomized.

Another issue in applying bootstrapping to this case pertains to how a second bias can be mistakenly included
if we are not careful to keep track of repeated entries of the same baseline data within a group.
In this case, the problem results from averaging used
to suppress residual signal covariances in the case that such cross-products may
actually be the product of a baseline's data with itself.  The noise bias in these
auto-products is coupled into the average of the covariances, and biases the result of the
off-diagonal covariance removal in each baseline group.  In the cross-multiplication of baselines
between groups, these biases (which are positive) couple into the result.

The solution in this case is to exclude auto-products from the determination of the average covariance
among baseline cross-products.  Since only a few independent cross-products are necessary
to distinguish baseline-specific features from a residual signal covariance, this restriction
is relatively harmless.  However, it does require that there be at least two independent baselines
in each resampled baseline group.  By imposing this restriction, as well as the restriction to not
repeat baselines between two independent groups, we limit the sample space that bootstrapping
explores.  It is therefore important to check that there remain enough data permutations to adequately
estimate errors.  Even with these restrictions on the resampling of 28 redundant baselines,
the number of 
distinct permutations is much larger than the number of independent samples we have, so
the limiting factor in estimating errors will be the number of samples, not the number of bootstrap
re-samplings.

The mean and 2$\sigma$ errors shown in Figure \ref{fig:pk_k3pk} are derived 
from 100 bootstrap re-samplings, with error bars calculated to enclose 95\% of the bootstrap samples.
The errors we report do not include covariances between $k$-modes, which are not treated in bootstrapping.
However, as implied in \citep{dillon_et_al2013b}, if these modes are not added together in binning, then
the variances derived from bootstrapping are sufficient to characterize the errors.

\section{Results and Discussion}
\label{sec:results}

\subsection{Foreground Suppression and Noise Levels}

\begin{figure*}\centering
\includegraphics[width=1.5\columnwidth]{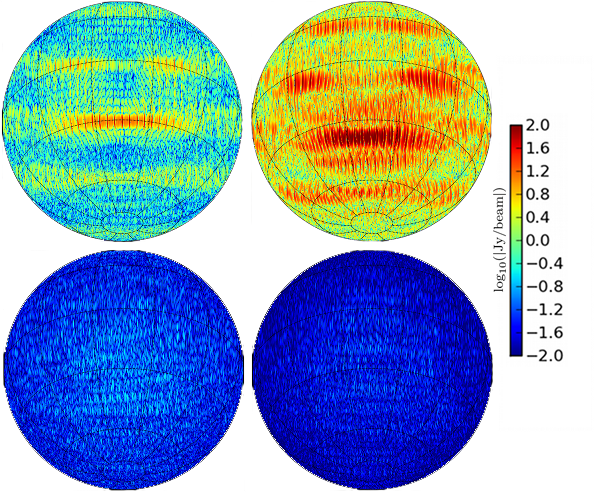}
\caption{
Synthesized beam and dirty images made with the same LST-averaged, maximum-redundancy
PAPER observations used in the power-spectrum analysis, but including all 496 baselines in the array.  
For imaging, we used every tenth integration (each with an integration time of 43s) 
in a single 500-kHz channel centered 
at 160 MHz in an LST range of 2:00 to 5:00,
phased to $\alpha$=3:00, $\delta$=-30:40$^\circ$.
Plotted left to right, top to bottom, are the synthesized beam, an
unfiltered dirty image, a dirty image with a wide-band delay filter at 15 ns past the horizon limit,
and a dirty image with both the wide-band delay filter and a fringe-rate filter 
applied.  The synthesized beam panel has
been normalized to show peak beam response at the maximum of the color scale range.
} \label{fig:imaging}
\end{figure*}

Figure \ref{fig:imaging} shows images of a single frequency channel using earth-rotation synthesis,
including all 496 baselines of PAPER's 32-antenna maximum-redundancy array.
As a reminder, the power spectra we report in \S\ref{sec:measured_pspec} is estimated
directly from a restricted set of visibilities.  These images are for diagnostic purposes.
As indicated by the marked reduction of emission between the upper-right and lower-left panels, the
application of the delay filter suppresses peak emission by at least 3 orders of magnitude.  The further decrease
in residual emission with the application of the fringe-rate filter between the lower-left and lower-right
panels indicates that residuals in the lower-left panel are noise-dominated, with the lower-right panel
showing that peak emission from the raw image has been suppressed by 3.3 orders of magnitude.

This result is paralleled in Figure \ref{fig:noise_vs_fq}, which shows the
RMS temperature of visibilities as a function of frequency in LST bins 
between 2:00 and 5:00, before and after the application of the
wide-band delay filter, as well as how noise in the filtered data behaves
through the various integration steps described in \S\ref{sec:analysis}.
As can be seen by the ratio
of the unfiltered data to the residual, the wide-band delay filter is responsible for 
suppressing foreground emission in the middle of the band by nearly three orders of magnitude (in mK),
demonstrating the efficacy of this approach for isolating smooth-spectrum
foregrounds from $k$-modes that might be used to detect 21cm reionization. This filter
substantially outperforms 
time-differencing
or frequency-differencing the visibilities
for foreground suppression
because of the sinusoidal nature of the fringes.

The gap in the lower set of curves of Figure \ref{fig:noise_vs_fq} between the raw and differenced
visibilities indicates the presence of an interfering signal that varies relatively smoothly in time and frequency.
Such characteristics are generally indicative of foreground emission, and suggest that near the limits
of the sensitivity of these observations, foregrounds are not completely suppressed.  
Examining the set of visibilities in Figure \ref{fig:noise_vs_fq} from which foregrounds have been removed
via derivatives in time or frequency,
we find a ratio of approximately 5 between the single-baseline and baseline-averaged curves in
the 140--180 MHz frequency range.  This ratio is roughly consistent with the value of 5.3 that 
would be expected for integrating 28 baselines with independent noise.  Below 130 MHz, this ratio decreases
substantially, indicating the presence of a correlated signal between different baselines.  This correlated
signal is shown to be suppressed by time-differencing and frequency-differencing filters, indicating
that it is likely to be the result of unsuppressed foregrounds near the minimum delay threshold imposed by the 
wide-band delay filter.  This interpretation is consistent with the power spectrum reported in Figure \ref{fig:pk_k3pk}.

Scaling the noise-dominated single-baseline curve by the 
square-root of the number of measurements contributing to each
temperature residual,
% sqrt(2 * 1e8 Hz / 203 ch * 42.8 s/day * 20 days * 2 pol), updated to reflect new PicA cal
we determine the system temperature at 160 MHz to be 560 K.  This value is in excess of the
expected level of sky noise arising from galactic synchrotron emission, but represents a substantial
improvement over the
system temperature in \citet{parsons_et_al2010}.

Comparing the set of visibilities derived from data after the application of 
a fringe-rate filter, which, although weighted
for a sharp cutoff in fringe-rate domain, effectively corresponds to a tenfold increase 
in integration time over the 43-second LST bins used in the upper plot.
The ratio between the curves with and without fringe-rate filtering is approximately 3 in the
140-170 MHz frequency range.  This ratio is roughly consistent with the value of 3.2 that would be
expected for the tenfold increase in integration time that the application of the fringe-rate filter represents.
The ratio between the filtered and unfiltered curves is substantially lower on either end of this band
indicating a departure from noise-like behavior.  As in other cases, the suppression of the systematic with
frequency-differencing suggests that this reflects residual foregrounds that are relatively smooth.
Nonetheless, the increasing systematics near the edge of the band form the basis of one of the arguments
for observing with a wide instantaneous bandwidth.  
It also provides the motivation for limiting our power spectrum analysis in this initial investigation
to the portion of the spectrum that appears consistent with noise.

\subsection{Measured Power Spectrum}
\label{sec:measured_pspec}

In Figure \ref{fig:pk_k3pk}, we show the spherically averaged 3D power spectrum derived
from our measurements.  In this figure, the central five data points in the
range $-0.06\le k_\parallel \le 0.06\ h\ {\rm Mpc}^{-1}$ fall within the horizon
limit of the delay spectrum, and come from data without the application
of the wide-band delay filter.  We also illustrate the data before (cyan) and
after (black) the application of the off-diagonal covariance removal described in \S\ref{sec:cov_diag}.
Averaging over $\pm k_\parallel$,
we place a 2$\sigma$
upper limit on $\Delta_{21}^2(k)$ of 1660 mK$^2$
for $k\approx0.27\ h\ {\rm Mpc}^{-1}$.  This limit is more than an order
of magnitude more stringent than the previous best upper limit, which was recently revised
to 61,500 mK$^2$ at $k=0.5\ h\ {\rm Mpc}^{-1}$ at $z=8.6$ \citep{paciga_et_al2013}.

%Outside of this range, measurements at higher
%$|k_\parallel|$ have significant noise components, but measurements in the range
%$-0.23\le k_\parallel\le0.16\ h\ {\rm Mpc}^{-1}$ do not, with a significance
%ranging upward from 2.7$\sigma$.  

As this plot illustrates, the residual foreground signal that we detect
in the frequency domain (see Figure \ref{fig:noise_vs_fq})
is concentrated at low $|k_\parallel|$-modes in delay space, with the largest
excess in a region just beyond the horizon limit described in P12b.  In the range 
$-0.22\le k_\parallel\le0.22\ h\ {\rm Mpc}^{-1}$,
bins show evidence of a systematic bias that makes them inconsistent with zero with high
significance.  The source of this biasing signal is likely to be foreground emission, as predicted in P12b, but
instrumental systematics are another possibility.
We observe a second feature at moderate (3$\sigma$) significance at $k_\parallel\approx\pm0.35\ h\ {\rm Mpc}^{-1}$
that appears relatively symmetrical with respect to positive and negative $k_\parallel$.  We presume
that this feature is caused by instrumental systematics or 
foreground emission, such as
leakage from a polarized signal with high rotation measure \citep{moore_et_al2013}.
The underlying cause is still under investigation, and 
will require substantial follow-up observation and careful analysis to validate.

As a diagnostic, we apply our analysis to random noise injected into the analysis before
the narrowband delay transform (see Figure \ref{fig:data_flow}) at the level expected
for these observations with a system temperature of 560 K, which was measured at 160 MHz using 
frequency-domain visibilities after the application of a wideband delay filter (Figure \ref{fig:noise_vs_fq}).  
These noise simulations
agree with Equation 27 of \citet{parsons_et_al2012a}, which is the basis of the dashed cyan curve
in Figure \ref{fig:pk_k3pk}.  We find good agreement between this expected noise level, the estimated
power spectrum, and the errors derived from bootstrapping.

\subsection{Constraints on the Mean 21cm Brightness Temperature}
\label{sec:brightness_bound}

\begin{figure*}\centering
\includegraphics[width=1.85\columnwidth]{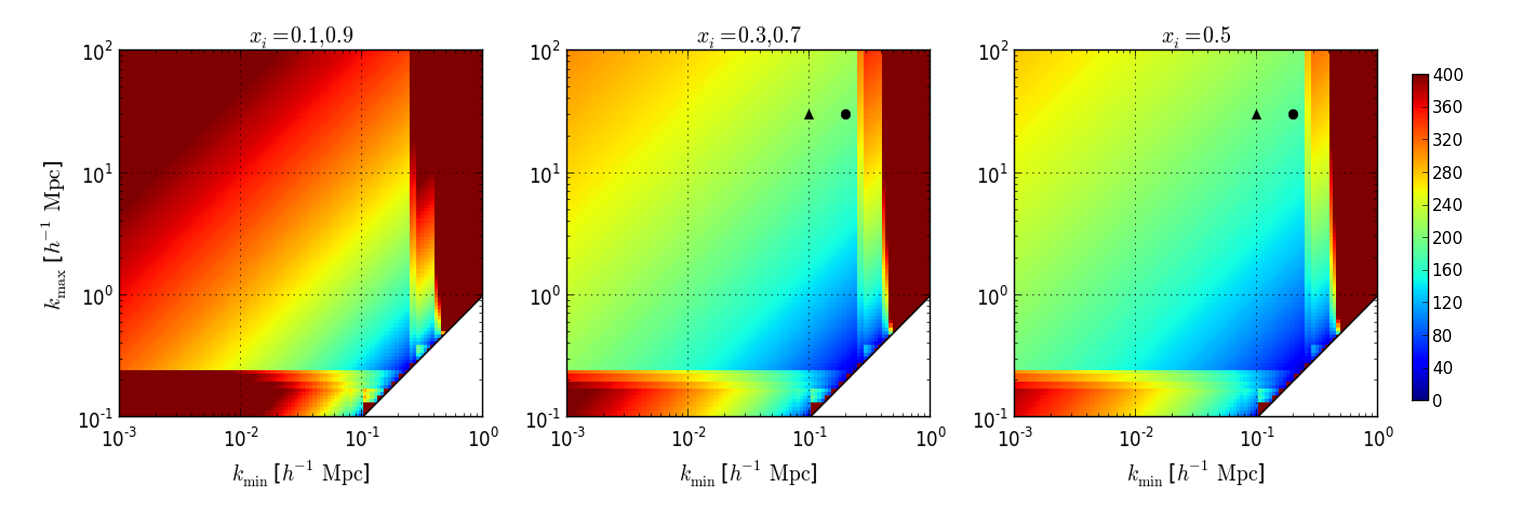}
\caption{
Constraints at $z=7.7$ on the absolute value of the 21cm brightness temperature of
the neutral gas (not normalized by ionization fraction), $|\tb|$, 
for the patchy reionization
model described in Equations \ref{eq:tune_tb} and \ref{eq:patchy_model}, as a function of 
the upper ($k_{\rm max}$) and lower ($k_{\rm min}$) bounds on the scale of fluctuations.
Color indicates the maximum $|\tb|$ (in mK) consistent with PAPER's 2$\sigma$ upper limits
on $\Delta^2_{21}(k)$ (see Figure \ref{fig:pk_k3pk}) for the listed ionization fractions.
Because the power spectrum amplitude for patchy reionization is invariant
under the interchange of ionized and neutral regions, $x_i=0.1$ and 0.3 equivalently correspond to
$x_i=0.9$ and 0.7, respectively.  The maximum color scale of 400 mK indicates the threshold of
brightness temperatures that
can be excluded a priori on the basis of the maximum contrast between the 21cm spin temperature
and the CMB (see Equation \ref{eq:MaxContrast}).  The black dot and black triangle indicate the coordinates
of a patchy-reionization approximation to the fiducial and high-mass halos models illustrated
in Figure \ref{fig:tb_limits}, respectively.
} \label{fig:patchy}
\end{figure*}

\begin{figure*}\centering
\includegraphics[width=1.85\columnwidth]{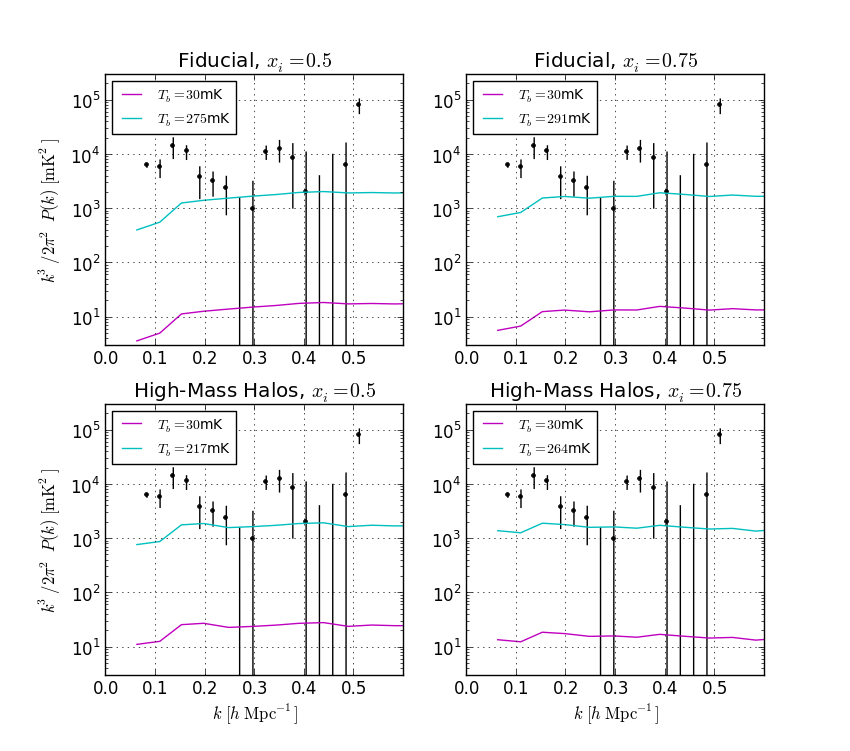}
\caption{
Scaled 21cm EoR power spectra predicted for the models described in
\S\ref{sec:brightness_bound} \citep{lidz_et_al2008},
reflecting bounds on the 21cm brightness temperature $\tb$ at $z=7.7$, according to
Equation \ref{eq:tune_tb}.  Magenta curves illustrate power spectra scaled according to
$\tb\approx30\ {\rm mK}$, as predicted by simulations that include the effects
of X-ray heating on the IGM.  Cyan curves illustrate power spectra scaled by the maximum
$\tb$ that is under the 2$\sigma$ upper limits we measure (black).  These are, left to right,
top to bottom, $\tb=275$, 291, 217, and 264 mK, respectively.
} \label{fig:tb_limits}
%# xi = .54, .71
%files_norm = [glob.glob('lidz_mcquinn_k3pk/power*%s*.dat' % s)[0] for s in ['7.3','7.0']]
%# xi = .51, .77
%files_hmass = [glob.glob('lidz_mcquinn_k3pk/hmass/power*%s*.dat' % s)[0] for s in ['7.4','7.1']]
%Reading lidz_mcquinn_k3pk/power_21cm_z7.32.dat c-
%Reading lidz_mcquinn_k3pk/power_21cm_z7.03.dat c--
%Reading lidz_mcquinn_k3pk/hmass/power_21cm_hmass_z7.47_a.dat m-
%Reading lidz_mcquinn_k3pk/hmass/power_21cm_hmass_z7.17_a.dat m--
\end{figure*}

The limits shown in Figure \ref{fig:pk_k3pk} are not yet at the level of the brighter predictions
favored by numerical and semi-numerical simulations
\citep{mcquinn_et_al2007,zahn_et_al2007,iliev_et_al2008,santos_et_al2008,mesinger_et_al2011} that
include, among other things, the predicted effects of X-ray heating on the IGM.  
Nonetheless, we can, for
the first time\footnote{The constraint in \citet{paciga_et_al2011} on the brightness temperature
of 21cm emission for a single-bubble reionization model was later retracted in \citet{paciga_et_al2013}.},
constrain the 21cm brightness temperature, $\tb$, of the neutral gas at mean density for
several reionization models and ionization fractions.  %Hence, we have suggestive
%evidence that 
%by $z=7.7$, the HI has been warmed 
%from its cold primordial state, probably by X-rays from mini-quasars or high-mass 
%X-ray binaries.
%\footnote{We define $T_b$ to
%be the mean 21cm brightness temperature of the neutral gas to indicate that it is not normalized by
%ionization fraction, as it is in the context of the global 21cm reionization signal in
%\citealt{furlanetto_et_al2006} and \citealt{pritchard_loeb2010}} inside of the $\sim500$ mK bound
%imposed by the maximum contrast between the CMB and the spin temperature, if it were coupled to
%the gas temperature assuming adiabatic expansion of the gas following recombination without heating.

% check this paragraph for stuff not in james paste
%In order to translate $k$-dependent upper bounds on $\Delta^2_{21}(k)$ into
%a constraint on $T_b$, we use the following prescription for relating the ionization 
%power spectrum $\Delta^2_i(k)$ that can be calculated from analytic or numerical models, to the predicted
%signal level in mK$^2$:
%\begin{equation}
%\Delta^2_{21}(k) = T_b^2 \Delta^2_i(k).
%\label{eq:tune_tb}
%\end{equation}
%To obtain the bounds on $T_b$ indicated in Figure \ref{fig:bubble}, we examine three different
%models for $\Delta^2_i(k)$ at various ionization states, and determine the 
%maximum $T_b$, beyond which the predicted value for
%$\Delta^2_{21}(k)$ is inconsistent with the measurements shown in Figure \ref{fig:pk_k3pk} at the 2$\sigma$ level.

% start james paste
We recall that the brightness temperature contrast of the 21cm transition
relative to the CMB (neglecting velocity gradient terms, and assuming low
optical depth, $\tau_{21}$) can be written in the form
\begin{align}
\label{eq:BrightnessFluctuations}
 \Delta T &\approx \frac{T_S - T_{CMB}}{1 +z} \tau_{21} \nonumber\\
 &\approx T_0(z) \left( \frac{T_S - T_{CMB}}{T_S} \right) x_{HI} (1 + \delta) 
\end{align}
where $T_S$ is the spin temperature, $x_{HI}$ is neutral fraction, and $\delta$
is the density contrast, all of which are spatially dependent. The
pre-factor is then
\begin{align}
 T_0(z) & = 23~{\rm mK}~\left( \frac{\Omega_b h^2}{0.02} \right) \left[ \left(
 \frac{0.15}{\Omega_m h^2} \right) \left( \frac{1+z}{10} \right) \right]^{1/2}\nonumber\\
 & = 25~{\rm mK}~\left( \frac{1+z}{10} \right)^{1/2}
\label{eq:Prefactor}
\end{align}
\citep[see, e.g., ][]{zaldarriaga_et_al2004,furlanetto_et_al2006}, where we have used the
Planck 2013 parameters \citep{planck_et_al2013}.  

Most reionization models assume an early
period of UV-photon production from stars that, while insufficient to ionize the
IGM, serves to drive the spin temperature towards the gas temperature via the
Wouthuysen-Field effect.  This is followed by heating of the neutral gas,
primarily through X-rays associated with star formation and black hole
accretion.  In these scenarios, heating drives $T_S \gg T_{CMB}$ during
reionization, so that $\xi \equiv(T_S-T_{CMB})/T_S \approx 1$.  The brightness
contrast can be significantly larger, however, if  $T_S < T_{CMB}$.
In this case, the 21cm line is seen in absorption against the 
CMB.

The maximum value of this contrast corresponds physically to a
situation in which the spin temperature remains tightly coupled to the gas
temperature via the Wouthuysen-Field effect\footnote{Recent measurements
\citep{bouwens_et_al2010,schenker_et_al2012} show that there are certainly enough UV photons for
this to be reasonable approximation at $z\sim8$.}
\citep{wouthuysen1952,field1958,hirata2006}, 
but in which heating of the neutral gas by any mechanism (X-rays, Ly-$\alpha$,
or shocks) is negligible.
While a full characterization of this ``no-heating'' scenario would require a
detailed simulation, we note that a reasonable re-scaling of power spectra for
which $\xi$ was assumed to be one (as in \citealt{lidz_et_al2008}) can be obtained in the
following way.  
Because reionization tends to rapidly produce either
fully neutral or fully ionized regions, $x_{HI}$ can be locally represented as either 0
(for an ionized region) or 1 (for a neutral one).  Since we have assumed no
heating of the neutral medium, the spatial dependence of $\xi$ is negligible.
This turns the product $T_0 \xi$ into an overall multiplicative factor on the
temperature fluctuations in Equation \ref{eq:BrightnessFluctuations}, and thus
re-scales the dimensionless ionization power spectrum, $\Delta^2_i(k)$
(calculated from analytic or numerical models), as
\begin{equation}
\Delta^2_{21}(k) = T_0^2(z) \xi^2(z) \Delta^2_i(k) \equiv \tb^2 \Delta^2_i(k),
\label{eq:tune_tb}
\end{equation}
We can estimate the maximum $\tb$ produced in the no-heating
scenario by noting that the CMB and 21cm spin temperature must stay in
equilibrium due to residual Compton scattering until about $z\sim 150$
\citep{furlanetto_et_al2006}.  Maximum contrast occurs if $T_S$
tracks the gas temperature thereafter, so that we can parameterize the maximum
value of $\xi$ as
\begin{equation}
 \xi_{\rm max} = 1 - \frac{1+z_{\rm dec}}{1 + z} \approx - \frac{150}{1+z},
\label{eq:Contrast}
\end{equation}
where $z_{\rm dec}$ is the redshift at which the kinetic temperature of the IGM
gas and the CMB drop out of equilibrium.  
Combining Equation \ref{eq:Contrast} and Equation \ref{eq:Prefactor} leads to
\begin{equation}
\label{eq:MaxContrast}
 \max[T_0(z) \xi(z)] \approx 370~{\rm mK}~\left( \frac{10}{1+z} \right)^{1/2}.
\end{equation}

To obtain a bound on $\tb$, we treat $\xi$ as a free parameter
ranging from $\xi_{\rm max}\le\xi\le1$ that parametrizes the
relative degree of heating of the IGM.  We
first note that brightness temperature
fluctuations in excess of Equation \ref{eq:MaxContrast} are ruled out
a priori at a given redshift, as indicated by the maximum color scale corresponding
to $\tb>400\ {\rm mK}$ in Figure
\ref{fig:patchy}.  We then examine three different models for $\Delta^2_i(k)$ at
various ionization states, and determine for each the maximum $\tb$, beyond which the
predicted value for $\Delta^2_{21}(k)$ is inconsistent at the 2$\sigma$ level
with the measurements shown in Figure \ref{fig:pk_k3pk}.
% end james paste

We examine two models presented in \citet{lidz_et_al2008} based on 
simulations by \citet{mcquinn_et_al2007}.  
Although the ionization power
spectra are associated with suggested redshifts in that paper, they apply to any redshift for
the indicated ionization fraction, $x_i$.  
The first model that we examine is the fiducial ``S1'' model.
For the $x_i=0.5$ and $x_i=0.75$ ionization power spectra, we obtain
limits of $\tb<275$ mK and $\tb<291$ mK, respectively.  These limits are illustrated in the upper
two panels in Figure \ref{fig:tb_limits}.

In the second model that we examine, ``S3'', reionization is
dominated by emission from massive halos, and
the 21cm power spectrum peaks at lower $k$-modes in the final stages of reionization because of
the increased spacing between the dominant sources of ionizing photons.  For this model, we obtain
limits of $\tb<217$ mK and $\tb<264$ for $x_i=0.5$ and $0.75$, respectively.  These limits are illustrated
in the lower two panels in Figure \ref{fig:tb_limits}.

The third model that we consider 
is a toy ``patchy''
reionization model described by Equation 21 in P12a, which approximates $\Delta^2_{21}(k)$ as being
flat between the bounds of a minimum ($k_{\rm min}$) and maximum ($k_{\rm max}$) cutoff in $k$, scaled
appropriately so that total power is conserved:
\begin{equation}
\Delta^2_i(k) = (x_{HI}-x_{HI}^2) / \ln(k_{\rm max}/k_{\rm min}),
\label{eq:patchy_model}
\end{equation}
where $x_{HI}=1-x_i$ is the neutral hydrogen fraction.
Using this toy model, we tune $k_{\rm min}$ and $k_{\rm max}$ and determine a maximum
mean temperature, above which Equation \ref{eq:tune_tb} becomes inconsistent with our measurements,
as shown in Figure \ref{fig:patchy}.
While our measurements are not yet able to exclude 
$\tb\approx 30\ {\rm mK}$ predicted for an X-ray-heated IGM under the reasonable assumption
that $k_{\rm max}/k_{\rm min}>10$,
we are able rule out $\tb\ge370 {\rm mK}$ for $k_{\rm min}<0.1\ h\ {\rm Mpc}^{-1}$ for nearly all physically
motivated values of $k_{\rm max}$ in both the $x_i=0.3$ and 
$x_i=0.5$ models.  For $x_i=0.1$, this result becomes dependent on the value of $k_{\rm max}$.

\subsection{Implications for X-ray Heating}

The constraints shown in Figures \ref{fig:patchy} and \ref{fig:tb_limits} indicate that, for several reionization models
and ionization fractions, our measurements are inconsistent with the 21cm brightness temperature from
neutral regions
that would result from the gas in the IGM cooling strictly according to adiabatic expansion
after recombination, without additional heating.  With the caveats that 1) the models we use 
may not correspond to the actual ionization power spectrum of reionization, 
and 2) we do not know whether the ionization fraction at $z\approx8$ is near to the
ionization fractions we investigate (although models such as \citealt{zahn_et_al2007} and
\citealt{lidz_et_al2008} do predict
$x_i=0.5$ near $z\approx8$), the implication of our measurements is that the IGM may have been heated with respect
to adiabatic cooling.  X-ray heating from inverse Compton scattering from supernovae,
high-mass X-ray binaries, or mini-quasars is the most likely culprit \citep{madau_et_al2004,ricotti_ostriker2004,mirabel_et_al2011,tanaka_et_al2012}, although shock heating, which
is currently disfavored \citep{mcquinn_oleary2012},  has not been completely ruled out as
a possibility.

We have used the PAPER results to explore some basic models for
reionization. Under the simplifying, although perhaps not unreasonable,
assumption that $\tb$ is roughly factorable from the growth of
structure (e.g. ionization bubbles and/or cosmic density), we constrain
$\tb$ in Figures \ref{fig:patchy} and \ref{fig:tb_limits} using limits to brightness temperature fluctuations
observed by PAPER.  For example, assuming a typical minimum $k$-scale of ionization fluctuations of
$0.1\ h\ {\rm Mpc}^{-1}$, and a maximum of $10\ h\ {\rm Mpc}^{-1}$, then as shown in
Figure \ref{fig:patchy}, the PAPER results imply upper limits on
$|\tb|$ of 191 and 175 mK for neutral fractions of 30\% and 50\%, respectively,
at $z=7.7$.  For currently favored reionization models
dominated by emission from massive halos \citep{mcquinn_oleary2012}, PAPER results
imply upper limits between 217 and 264 mK.

In summary, for massive-halo-dominated reionization models, and for a range of simple 
patchy reionization models, PAPER data imply that the
value of $\tb$ is less than 400 mK.  The value of 400 mK is
important, since it reflects the expected $\tb$ in the case where the
spin temperature of the 21cm line is coupled to the gas kinetic temperature,
but the neutral gas is never heated on large scales due to, e.g. X-rays
from high-mass X-ray binaries or mini-quasars.  Physically, this could correspond
to a situation where the Ly$\alpha$ photons from the first galaxies are
sufficient to couple the spin and kinetic temperatures throughout the neutral IGM
via the Wouthuysen-Field effect,
but they do not substantially heat the gas \citep{chen_et_al2004}, nor do
supernovae, X-ray binaries, or mini-quasars produce enough X-rays to heat the large scale neutral
IGM.
We point out that while the X-ray heating constraints we present require a coupling between the spin
and kinetic temperatures, we know that this coupling must be occurring at $z=7.7$. 
Observed star formation rates (SFRs) at $z=8$ \citep{bouwens_et_al2010,schenker_et_al2012}
exceed by almost an order of magnitude
the threshold of $10^{-3} M_\odot/{\rm yr}\ {\rm Mpc}^3$ required for UV photons to
couple these temperatures \citep{mcquinn_oleary2012}.

While this no-heating model is considered physically unlikely \citep{furlanetto_et_al2006},
the argument can be turned around: the fact that we constrain
fluctuations at the 400 mK level provides, for the first time, suggestive evidence
for large-scale heating of the neutral IGM during reionization at $z>7$.
Such constraints on X-ray heating may soon have repercussions on the global signal
\citep{pritchard_loeb2012} and on the morphology of ionization \citep{mesinger_et_al2013}.
Note that our measurements pertain to the thermal evolution of the
neutral gas during reionization, and not the subsequent thermal
evolution of the ionized gas after reionization \citep{hui_haiman2003}.

\section{Conclusion}
\label{sec:conclusion}

We have set the most stringent limits to date on the 21cm EoR power spectrum.  This achievement
was greatly facilitated by
using the delay-spectrum approach to achieve a level of foreground
isolation of -80 dB outside of the horizon limit.  This is the 
deepest level of foreground suppression that has been achieved so far by a 21cm reionization experiment,
and with a $2\sigma$ upper limit of $\Delta^2_{21}(k)<(41 {\rm mK})^2$ at $k=0.27\ h\ {\rm Mpc}^{-1}$,
the result is within an order of magnitude (in mK) of predictions of the 21cm 
EoR signal level.  These constraints are sufficient to begin ruling out several scenarios
where X-ray and shock heating fail to heat the IGM prior to reionization, although we cannot rule out
the possibility of a cold IGM with low neutral fraction at $z=7.7$.

Although PAPER has less collecting area than other competing 21cm reionization
experiments, its ability to access lower $k$-modes than, e.g., \citet{paciga_et_al2011}, as
a result of its smooth instrumental responses, combined with the sensitivity boost that arises from
the use of a maximum-redundancy array configuration, 
shows the effectiveness of the approach taken by PAPER.
The efficacy of the wide-band delay filter in this analysis shows PAPER's wide operational
bandwidth to be a valuable asset for isolating foreground emission in $k$-space.
This work represents an important first validation of the delay-spectrum approach to
avoiding foregrounds.

At the limit of the sensitivity of these observations, we are beginning to see signs of bias at
a range of $k$-modes,
presumably arising from foregrounds or other systematics.  Additional investigation
will be necessary to further diagnose
the source of this bias and constrain its source.  
Using additional data that have been observed but not yet analyzed, and a novel filtering
technique, we expect forthcoming analysis to yield nearly an
order-of-magnitude improvement in sensitivity, in units of mK$^2$, over the results presented here.
Additionally, we also intend to explore this power-spectrum analysis over a wider bandwidth
to constraint a broader range of cosmological history.
Meanwhile, observations have completed with a 64-antenna, dual-polarization PAPER deployment in
South Africa, and a new observing campaign is just beginning (as of late 2013) with a 128-antenna 
deployment of PAPER at the same site.
Plans are also proceeding for a
next-generation instrument, the Hydrogen Epoch of Reionization Array (HERA)\footnote{\url{http://reionization.org/}}, which, with $\sim0.1 {\rm km}^2$ of collecting area,
will have the capability
to characterize the power spectrum of reionization in detail \citep{pober_et_al2013b}.

\section{Acknowledgment}

We would like to thank SKA-SA for the site infrastructure, maintenance, and observing support
that has made this work possible, as well as the significant efforts of the staff at
NRAO's Green Bank and Charlottesville sites.  AP would like to thank M. McQuinn and A. Lidz
for helpful discussions and ionization models.
The PAPER project is supported
by the National Science Foundation (awards 0804508,
1129258, and 1125558), and a generous grant
from the Mt. Cuba Astronomical Association.

% ---------------------------------------------------------------------
% ---------------------------------------------------------------------
% ---------------------------------------------------------------------

\appendix
\section{A Data Compression Technique for Low-Frequency Telescopes}
\label{app:data_compression}

Managing the volume of data generated by modern radio telescopes is a technical challenge that is becoming
increasingly problematic. 
For example, the volume of raw data on which this paper's results are based exceeds 10 Terabytes.  With
data volume scaling quadratically with antenna number, PAPER observations in the near future are
likely to occupy more than 0.5 Petabytes.  This problem is common to many current arrays,
and several data-reduction schemes (e.g. \citealt{ord_et_al2010}) have been proposed.
These schemes generally place a substantial burden on real-time calibration, a feat that, while
possible, often proves elusive, particularly for arrays that are under active development.

In this section, we describe a new method of (lossy) data compression based on
delay/delay-rate (DDR) filters \citep{parsons_backer2009} that
are tailored to the geometric limits of celestial emission in an
interferometer's native observing coordinates, omitting data outside of those limits.
In contrast to data-reduction pipelines
that require real-time calibration, DDR-based compression requires a minimal amount
of information at the time of observation, provided that instrumental responses
are smooth in time and frequency.  In the minimal limit, substantial compression
proceeds based simply on knowing the maximum baseline length in the array.  

\begin{figure*}\centering
\includegraphics[width=0.75\columnwidth]{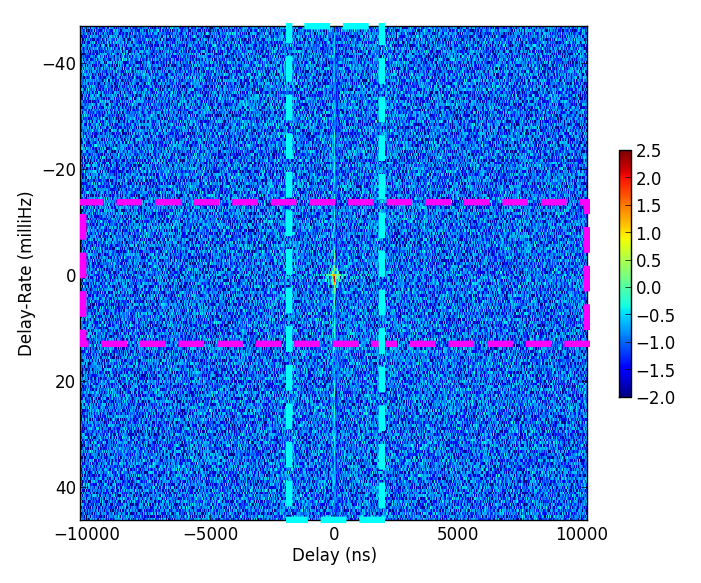}
\caption{
The delay/delay-rate (D/DR) transform of RFI-flagged visibilities from 30-m PAPER baselines.  Raw visibilities
are integrated for 10.7 s with a channel bandwidth of 50 kHz to facilitate the detection and removal of
RFI and other transient events.  The regions of D/DR-space sampled according to these observing parameters far
exceed the delays (cyan) and delay-rates (magenta) that can be occupied by sky emission that smoothly varies
versus time and frequency, as dictated by a maximum baseline length of 300 m in the array.  Color scale indicates
$\log_{10}({\rm Jy})$, with arbitrary scaling.  Although D/DR filtering is applied to
each baseline individually, redundant 30-m baselines were co-added in this figure to
suppress noise and better illustrate the regions occupied by sky signal.
} \label{fig:ddr_compression}
\end{figure*}

To briefly review the principles of DDR filters presented in \citet{parsons_backer2009},
the geometric delay with which a celestial signal originating in direction $\hat s$ enters
an interferometric baseline $\b$ is given by
\begin{equation}
\tau_g=\b\cdot \hat s / c
\end{equation}
where $c$ is the speed of light.
Following directly from this definition, the bounds on the geometric delay for this baseline are
\begin{equation}
-\frac{|\b|}{c}\le\tau_g\le\frac{|\b|}{c}.
\label{eq:horizon_bound}
\end{equation}
For the maximum baseline length of 300 m, $|\b|/c$ is approximately 1$\mu$s,
which corresponds to
$-0.52\le k_\parallel<0.52\ h\ {\rm Mpc}^{-1}$ at $z\approx8$.
An example of the bounds of a delay filter in DDR space is given
by the shaded cyan region in Figure \ref{fig:ddr_compression}.
It is worth noting that the range of $k_\parallel$ that are occupied by foregrounds for
the 30-m baselines used in the power spectrum analysis is substantially narrower than
the maximum bounds placed by this data compression step.  This leaves a substantial region
for probing the cosmological 21cm signal that includes the modes to which we are most sensitive.

Taking the Fourier transform of
a visibility along the frequency axis (see Equation \ref{eq:dtransform}) produces a 
delay-domain expression of the visibility, which 
we can approximate for a model sky of point sources indexed by $n$ as
\begin{equation}
\Vt(\tau,t)=\sum_n{\At(\tau,\hat s_n(t)) * \tilde{S}_n(\tau) * \delta_{\rm D}(\tau_{g,n} - \tau)},
\end{equation}
where $\At(\tau,\hat s_n(t))$ indicates the Fourier transform of the antenna response in source direction $\hat s_n$ along the
frequency axis, $\tilde{S}_n$ indicates the Fourier transform of the source flux density along the frequency
axis, $\delta_{\rm D}$ is a delta function relating delay to the geometric delay $\tau_{g,n}$ in the
direction of the source indexed by $n$, and `$*$' indicates a convolution in delay.

Provided that celestial emission and
instrumental responses are smooth versus frequency, so that $\At$ and $\tilde{S}$ are compact in $\tau$, 
emission in $\Vt$ will be tightly bounded by the geometric constraints on $\tau_g$ given
in Equation \ref{eq:horizon_bound}, and a low-pass filter may be
applied to $\Vt$ that preserves this range of $\tau$-modes.  In doing so, such a filter preserves
smooth-spectrum celestial emission, and removes modes that are expected to consist mostly of noise.  Moreover,
with the range of delay modes now being band-limited, visibilities may be decimated in frequency without
aliasing, according to the critical Nyquist rate.  This decimation is the basis of
data compression in the frequency direction.

Similar geometric limits apply to the variation of visibilities in the time dimension.  As
described in Equation 7 of \citet{parsons_backer2009}, the rate of change of the geometric delay versus
time --- that is, delay-rate --- is given by
\begin{equation}
\frac{d\tau_g}{dt}=-\omega_\oplus\left(\frac{b_x}{c}\sin H + \frac{b_y}{c}\cos H\right)\cos\delta,
\end{equation}
where $\b=(b_x,b_y,b_z)$ is the baseline vector expressed in equatorial
coordinates, $\omega_\oplus$ is the angular frequency of the earth's rotation, and $H,\delta$ are the
hour-angle and declination of a point on the celestial sphere, respectively.  As a result, there exists a maximum
rate of change based on the length of a baseline projected to the $z=0$ equatorial plane.
For arrays not deployed near the poles, $|b_y|\gg|b_x|$ (i.e.,
they are oriented more along the east-west direction than radially from the polar axis),
and the maximum rate of change corresponds to $H=0$ and $\delta=0$, where we have
\begin{equation}
-\omega_\oplus\frac{|b_y|}{c}\le\frac{d\tau_g}{dt}\le\omega_\oplus\frac{|b_y|}{c}.
\end{equation}
For a maximum east-west baseline length in the PAPER array of 300m, $\omega_\oplus|b_y|/c$ is approximately
0.07 ns/s.  
To better elucidate the meaning of this bound, we take the Fourier transform 
along the time axis (see Equation 8 in \citealt{parsons_backer2009}) for a model visibility
consisting of a single point source located at the point of maximum delay-rate, which gives us
\begin{align}
\Vt(\nu,f)&\approx\At(\nu,f) * \tilde{S}(\nu) * \int{e^{2\pi i\omega_\oplus\frac{b_y\nu}{c}t}e^{-2\pi ift}dt}\nonumber\\
&\approx\At(\nu,f) * \tilde{S}(\nu) * \delta_{\rm D}(\frac{b_y}{c}\omega_\oplus \nu - f),
\end{align}
where $f$ is the fringe-rate of the 
source\footnote{Delay rate is equivalent to the frequency-integrated fringe rate.}, 
$\At(\nu,f)$ indicates the Fourier transform of the antenna response along the time direction,
and approximation is
indicated because we assume $|b_y|\gg|b_x|$, and because the Fourier transform must involve
a discrete length of time, during which our assumption that $\cos H\approx1$ breaks down at second order.
The delta function above gives rise to the expression for the maximum fringe rate in Equation \ref{eq:fringe_rate}.

This example of a source with a maximal fringe-rate serves to show that a 
filter may be applied in delay-rate domain, using the fact that the maximum
delay-rate is bounded by the maximum fringe rate within the band (i.e. evaluating Equation \ref{eq:fringe_rate}
at the maximum $\nu$ involved in the delay transform), to remove emission that exceeds the variation
dictated by array geometry for sources locked to the celestial sphere.  As in the delay filtering case, assuming
the geometric bounds on delay rate implicitly assumes that $\At$ and $\tilde{S}$ are compact in $f$, which
is to say that instrumental responses and celestial emission must be smooth in time; variable
emission from, e.g., fast-transients will be heavily suppressed by such delay-rate filters.
For PAPER, with a maximum baseline length of 300m and a maximum observing frequency of 200 MHz, 
the maximum delay-rate has a period of 68.5s.  As described in \S\ref{sec:fringe_rate_filtering}, at PAPER's
latitude, delay-rates range from -$f_{\rm max}/2$ to $f_{\rm max}$.  Filtering delay-rates to this
range corresponds in sensitivity to an effective integration time of 45.2 s.
An example of the bounds of a delay filter in DDR space is given
by the shaded magenta region in Figure \ref{fig:ddr_compression}.  As in the delay filtering case,
filtering along the delay-rate axis permits substantial down-sampling of the signal, which is
the basis for the reduction in data volume along the time axis.
We note that for the analysis
in \S\ref{sec:preprocessing}, we choose to use a slightly wider delay-rate filter to be conservative in
our first application of this technique, corresponding to an integration time of 43 seconds.

A serious issue associated with DDR filtering pertains to how data flagging associated with the
removal of RFI and systematics violates the assumptions of spectral and temporal smoothness
that are implicit in the geometric bounds we derive for delay and delay-rate.  This issue is
addressed in \citet{parsons_backer2009} by a CLEAN-like iterative forward modeling deconvolution in delay domain for removing the sidelobes
created by the unsmooth sampling functions associated with data flagging.  We apply the same
technique here for DDR-based data compression.  We also take advantage of the substantial
suppression of celestial emission that results from the removal of low delay and delay-rate modes to
improve RFI flagging.  Hence, our approach to implementing DDR-based data compression proceeds
as follows:
\begin{enumerate}
\item Coarse RFI flagging, using time and frequency derivatives to suppress celestial emission.
\item DDR transformation and CLEAN deconvolution of data from a subset of baselines.  This culminates 
in a model of emission for each of these baselines, with model components restricted to fall within 
specified geometric limits.  In order to avoid edge effects when filtering along the time axis, we filter
data windows that overlap by a factor of 3 and retain only the central third of the filtered data.
\item Refined RFI flagging of residuals after subtracting the above model from each baseline in the subset.
\item Application of the RFI flagging to all baselines on the basis of occupancy in time/frequency bins.
\item DDR transformation and CLEAN-like deconvolution of raw data from all baselines, based on the refined RFI flagging.
\item Filtering of DDR-domain data into regions that are saved to three separate output files, as follows:
\begin{enumerate}
\item Smooth-Spectrum Non-Transient Emission.  These data are filtered at the
intersection of the delay and delay-rate low-pass filters (cyan and magenta, respectively,
in Figure \ref{fig:ddr_compression}) and decimated to the critical Nyquist rate along
both time and frequency axes.  Data are saved to disk using a high dynamic range 32-bit floating point data type.
Compression factor for PAPER: $\sim$40.
\item Smooth-Spectrum Transient Emission.  These data are filtered by a
low-pass delay filter and a high-pass delay-rate filter (cyan in Figure \ref{fig:ddr_compression},
excluding the magenta region)
and decimated along the frequency axis to the
critical Nyquist rate.  Since the bulk of celestial emission has been suppressed by not including low-order
delay-rate modes, these data are written using a lower dynamic range 16-bit shared exponent
floating point data type that is adequate for representing a noise-like signal.
Compression factor for PAPER: $\sim$20.
\item Unsmooth Non-Transient Emission.  These data are filtered by a
high-pass delay filter combined with a low-pass delay-rate filter (magenta in Figure \ref{fig:ddr_compression},
excluding the cyan region) and decimated along the time axis to
the critical Nyquist rate.  These data are also written in the lower dynamic range data type described above.
Compression factor for PAPER: $\sim$8.
\end{enumerate}
\end{enumerate}

\begin{figure*}\centering
\includegraphics[width=0.75\columnwidth]{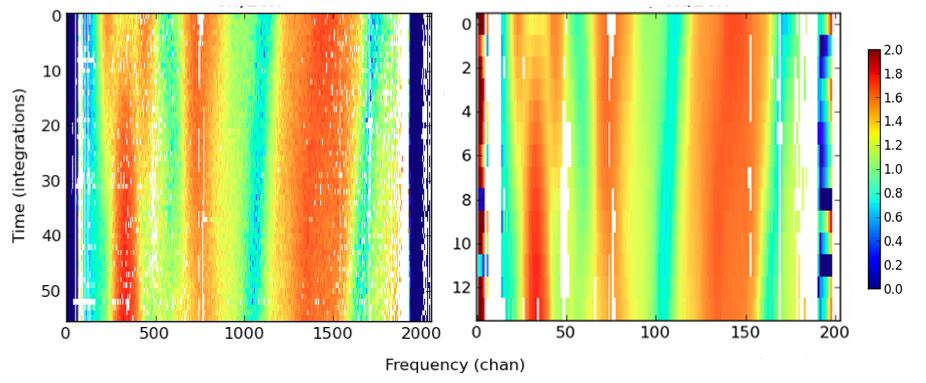}
\caption{
The dynamic spectrum of RFI-flagged visibilities from a 30-m PAPER baseline prior to (left) and after (right)
the application of data compression along the delay and delay-rate axes.  
Color scale indicates $\log_{10}({\rm Jy})$, with arbitrary scaling.
} \label{fig:compression_before_after}
\end{figure*}

The result of the application of this procedure for smooth-spectrum non-transient emission 
to data from one of the baselines used in this
analysis is shown in Figure \ref{fig:compression_before_after}.  As shown, the spectrally and temporally smooth
components of the signal are preserved by data compression, while the unsmooth components that
originate predominantly from thermal noise are removed.
This compression algorithm is lossy, but the regions lost in the compression
(namely, the unshaded corners in Figure \ref{fig:ddr_compression}) are
likely to be of minimal import for many low-frequency radio telescopes.

The computational cost of DDR compression is dominated by the deconvolution of
sampling function that is required because of RFI flagging. The CLEAN algorithm is nonlinear, but the
computation time generally scales as $N^2$, where $N$ is the number
of antennas in an array.  A 35-core cluster was sufficient to compress 32-antenna dual-polarization
PAPER observations roughly four times faster real time.  For an upcoming 128-antenna deployment
of PAPER, we will be deploying a 64-core cluster that
is expected to be sufficient to compress data two times faster than
real time.

This technique can be easily generalized to other
arrays, as it requires minimal information about the geometry of an array.  Compression factors could
be improved by more carefully tailoring DDR filters to the geometry of each baseline individually.
We have not explored this on PAPER, as the advantages of a compression system that requires minimal
information about the array at run-time and is minimally restrictive with respect to further
analysis seem to outweigh the need for additional data compression at this time.

\section{On Calculating Beam Areas for Normalizing Power-Spectrum Measurements}
\label{app:beam_area}

In this section, we investigate in greater detail the issue of how beam area enters into the power-spectrum
formalism for 21cm intensity mapping derived in \citet{morales2005}, \citet{mcquinn_et_al2006}, \citet{pen_et_al2009},
and P12a.  In particular,
we show that, for the purpose normalizing power spectrum measurements by the integrated volume,
the appropriate beam area to use is proportional to the integral of the square of the power
response.  This fact has been obscured in many derivations in the literature by certain approximations
that have been made to make volume integrals analytically tractable.  
This derivation represents the general form of the special case for a Gaussian beam that was correctly
derived in \citet{pen_et_al2009}.

We begin by examining the integrated volume, $\mathbb{V}$, used to normalize the 3D Fourier transform in
Equation 3 of P12a. We express this volume in observing coordinates as
\begin{equation}
    \mathbb{V} = \Omega B\cdot X^2Y,
\end{equation}
where $B$ is the bandwidth, $\Omega$ is the angular area, and $X,Y$ are redshift-dependent
scalars relating angle and frequency to spatial scales, respectively.
$\Omega$ arises from the bounds set by $A(l,m,\nu)$, the antenna power response, on the 
angular extent in the integral
\begin{align}
    \Vt^2(u,v,\eta)=&\left(\frac{2k_{\rm B}}{\lambda^2}\right)^2
        \left[\int{dl~dm~d\nu A(l,m,\nu)T(l,m,\nu)e^{-2\pi i(ul+vm+\eta\nu)}}\right]\times\nonumber\\
        &\left[\int{dl^\prime~dm^\prime~d\nu^\prime A^*(l^\prime,m^\prime,\nu^\prime)T^*(l^\prime,m^\prime,\nu^\prime)e^{2\pi i(ul^\prime+vm^\prime+\eta\nu^\prime)}}\right],
\end{align}
which is a slightly modified version of Equation 6 of P12a relating the delay-transformed visibility
$\Vt$, sampled at wavemodes $u,v$ (the Fourier complements of angular coordinates $l,m$) and $\eta$ (the Fourier
complement of spectral frequency $\nu$), to a temperature field $T$.  Re-expressing this in real-space
cosmological coordinates, we have
\begin{align}
    \Vt^2(u,v,\eta)=&\left(\frac{2k_{\rm B}}{\lambda^2}\right)^2
        \frac{1}{X^2Y}\left[\int{d^3r A(\r)T(\r)e^{-i\k\cdot\r}}\right]\times\nonumber\\
        &\frac{1}{X^2Y}\left[\int{d^3r^\prime A^*(\r^\prime)T^*(\r^\prime)e^{i\k\cdot\r^\prime}}\right].
\end{align}
We may then use the definition of the Fourier transform (following the convention of Equation 1 in P12a) to write
\begin{align}
    \Vt^2(u,v,\eta)=&\left(\frac{2k_{\rm B}}{\lambda^2}\right)^2\frac{\V^2}{(X^2Y)^2}
        \int{\frac{d^3k^\prime}{(2\pi)^3} \At(\k-\kp)\Tt(\kp)}\times\nonumber\\
        &\int{\frac{d^3k^{\prime\prime}}{(2\pi)^3} \At^*(\k-\kpp)\Tt^*(\kpp)},
\end{align}
with the multiplicative terms in the previous equation now appearing as convolutions, and $\At$ indicating
the 3D Fourier transform of the antenna power response.  Combining the
two integrals yields
\begin{align}
    \Vt^2(u,v,\eta)&=\left(\frac{2k_{\rm B}}{\lambda^2}\right)^2\frac{\V^2}{(X^2Y)^2}
        \int{\frac{d^3k^\prime}{(2\pi)^3}\frac{d^3k^{\prime\prime}}{(2\pi)^3}\At(\k-\kp)\At^*(\k-\kpp)\Tt(\kp)\Tt^*(\kpp)}\nonumber\\
    &=\left(\frac{2k_{\rm B}}{\lambda^2}\right)^2\frac{\V}{(X^2Y)^2}
        \int{\frac{d^3k^\prime}{(2\pi)^3}\left|\At(\k-\kp)\right|^2\widehat P(\kp)},
\end{align}
where the second step uses the relationship between an estimate of the power spectrum $\widehat P(\kp)$ 
and the two-point
correlation function of a temperature field given in Equations 2 and 3 in P12a.  As discussed in
P12a and \citet{mcquinn_et_al2006}, the narrow width of $\At$ in $k$-space in comparison to the
scales on which $\widehat P(\k)$ varies
allows $\widehat P(\k)$ to be removed from the integral, giving us
\begin{equation}
    \Vt^2(u,v,\eta)\approx\left(\frac{2k_{\rm B}}{\lambda^2}\right)^2\frac{\V}{(X^2Y)^2}
        \widehat P(\k)\int{\frac{d^3k^\prime}{(2\pi)^3}\left|\At(\k-\kp)\right|^2}.
\end{equation}
Using the coordinate substitution $\q\equiv\kp-\k$, we have
\begin{equation}
    \Vt^2(u,v,\eta)\approx\left(\frac{2k_{\rm B}}{\lambda^2}\right)^2\frac{\V}{(X^2Y)^2}
        \widehat P(\k)\int{\frac{d^3q}{(2\pi)^3}\left|\At(\q)\right|^2}.
\end{equation}
Finally, we invoke Parseval's theorem to substitute the integral of the variance in real-space,
picking up a factor of $1/\V$ in accordance with our chosen Fourier convention, so that we have
\begin{align}
    \Vt^2(u,v,\eta)&\approx\left(\frac{2k_{\rm B}}{\lambda^2}\right)^2\frac1{(X^2Y)^2}
        \widehat P(\k)\int{d^3r\left|A(\r)\right|^2}\nonumber\\
        &\approx\left(\frac{2k_{\rm B}}{\lambda^2}\right)^2\frac1{X^2Y}
        \widehat P(\k)\int{dl~dm~d\nu\left|A(l,m,\nu)\right|^2}\nonumber\\
        &\approx\left(\frac{2k_{\rm B}}{\lambda^2}\right)^2\frac{B}{X^2Y}
        \widehat P(\k)\int{dl~dm\left|A(l,m)\right|^2}.
\end{align}

We compare this result with the relation between the delay-transformed visibility,
${\tilde V}$, to the three-dimensional power spectrum of reionization, $P_{21}(\k)$ 
(P12a):
\begin{equation}
    {\tilde V}_{21}^2(u,v,\eta)\approx\left(\frac{2k_{\rm B}}{\lambda^2}\right)^2\frac{\Omega\,B}{X^2Y} \widehat P_{21}(\k).
    \label{eq:v2_vs_pk}
\end{equation}
As this shows, the relevant beam area in Equation \ref{eq:v2_vs_pk} is the
power-square beam, $\Omega_{\rm PP}$, given by
\begin{equation}
\Omega_{\rm PP}\equiv\int{dl~dm\left|A(l,m)\right|^2}.
\label{eq:beam_squared}
\end{equation}
This contrasts with the standard metric for beam area --- the integrated
power beam --- which we will call $\Omega_{\rm P}$, and is given by
\begin{equation}
\Omega_{\rm P}\equiv\int{dl~dm A(l,m)},
\label{eq:beam_area}
\end{equation}
This beam area metric is used to convert visibility measurements from Jy units to mK,
but is incorrect for normalizing power spectra that relate to 
the two-point correlation function of a temperature field.

In the literature to date, sensitivity derivations have commonly assumed a top-hat beam response
or a sinc response in the $uv$-plane (see, e.g., \citealt{morales2005}, \citealt{mcquinn_et_al2006}, and P12a).
Since $\Omega_{\rm P}=\Omega_{\rm PP}$ under these assumptions, this has permitted some ambiguity as to which
beam area is intended.  For equations that relate power-spectrum sensitivity to a system temperature
(e.g. Equations 15 and 16 in P12a)
\begin{equation}
\Omega^\prime\equiv\Omega_{\rm P}^2/\Omega_{\rm PP}
\end{equation}
should be used in lieu of $\Omega$, as
these equations pick up two factors of $\Omega_{\rm P}$ in the conversion from Jy$^2$ to mK$^2$, along with
the a factor of $\Omega_{\rm PP}$ in the denominator relating to the integrated volume.
For equations that relate a measured visibility (in units of brightness
temperature, e.g. Equation \ref{eq:pspec_cosmo}) to $\widehat P(\k)$, the factor of 
$\Omega_{\rm P}^2$ is already 
applied in the conversion from units of Jy to mK, 
and $\Omega$ corresponds to the remaining factor of $\Omega_{\rm PP}$. 

For PAPER, $\Omega_{\rm P}\approx0.72$ sr, while $\Omega_{\rm PP}$ is 0.31 sr.  Following the definition
above, $\Omega^\prime\approx1.69$.  These beam areas are calculated numerically from
a beam model, but typically, $\Omega^\prime$ is about a factor of two larger than $\Omega_{\rm P}$.

\section{Optimality of Analysis for Removing Off-Diagonal Covariances}
\label{app:cov_diag}

In this section, we investigate delay-domain analysis through the lens of the
quadratic estimator formalism \citep{liu_tegmark2011,dillon_et_al2013a},
ultimately showing that the analysis technique of removing off-diagonal covariances
arising from systematics
produces optimal power spectrum estimators, to leading order.

\subsection{Review of Quadratic Estimator Formalism}
We begin by reviewing the quadratic estimator formalism.
In this formalism, the value of the power
spectrum $p_\alpha$ in the $\alpha$th $k$-bin (i.e. $P(k_\alpha)$, also known as the ``bandpower") can be
optimally estimated by the quantity\footnote{Equation \ref{eq:estimator} in principle requires an extra bias removal term in order to be optimal \citep{liu_tegmark2011,dillon_et_al2013a}.  For simplicity we omit this term in anticipation of Section \S\ref{sec:DelayTransApp}, where we will generalize our results to cross-power spectra between baselines.  Since cross-power spectra do not incur noise biases \citep{dillon_et_al2013b}, our final result will not require the bias removal term.}
\begin{equation}
\label{eq:estimator}
\widehat{p}_\alpha = \frac{1}{2} \sum_\beta M_{\alpha\beta} \x^\dagger \C^{-1} \Q^\beta \C^{-1} \x,
\end{equation}
where the hat signifies that $\widehat{p}_\alpha$ is an estimate of the true
bandpower $p_\alpha$, $\x$ is a vector containing binned data with
$\x_\alpha\equiv \Vt(\tau_\alpha)$ from Equation \ref{eq:dtransform}, $\mathbf{C} \equiv
\langle \x \x^\dagger \rangle$ is the covariance matrix of the data,
$M_{\alpha\beta}$ (discussed below) normalizes the estimator, and $\mathbf{Q}^\beta$ is the
response of the covariance to the $\beta$th bandpower.  The family of $\Q$ matrices (one for each value of $\beta$) map the vector
space of bandpowers to the vector space of data, and is constructed to satisfy
the relation
\begin{equation}
\C^\textrm{sky} = \sum_\beta p_\beta \Q^\beta,
\end{equation}
where $\C^\textrm{sky}$ is the contribution of sky emission to the covariance.  This contribution includes both the foregrounds and cosmological signal.  The foregrounds are included with the signal because we have conservatively avoided the direct modeling (and subtraction) of foregrounds from the data \citep{dillon_et_al2013b}.  The total covariance is given by
\begin{equation}
\label{eq:Qdef}
\C = \N + \C^{\textrm{junk}} +  \C^{\textrm{sky}},
\end{equation}
where $\N$ is the instrumental noise covariance, and $\mathbf{C}^{\textrm{junk}}$ is
the contribution of all contaminants except for foregrounds and noise.

Thinking of $M_{\alpha\beta}$ as being an element of a matrix $\M$ that lives in a vector space of
bandpowers, one is free to choose between several different choices of $\M$,
all of which are lossless (since $\M$ is generally taken to be an invertible matrix, and
all steps prior to its application are also lossless) but have different error properties.  Consider for
example the window function matrix $\mathbf{W}$, which relates the expectation
value of our bandpower estimators to the true bandpowers:
\begin{equation}
\langle \mathbf{\widehat{p}} \rangle = \mathbf{W} \mathbf{p},
\end{equation}
where we have grouped the bandpowers into vectors.  The window function matrix can be shown to take the form
\begin{equation}
\mathbf{W} = \M \F,
\end{equation}
where $\F$ is the Fisher information matrix, given by
\begin{equation}
\label{eq:FishDef}
F_{\alpha \beta} = \frac{1}{2} \textrm{tr} \left[ \C^{-1} \Q^\alpha \C^{-1} \Q^\beta \right].
\end{equation}
Different choices for $\M$ give window functions (rows of $\mathbf{W}$) with
different properties, and one is subject only the restriction that $\M$ must be
normalized in a way that ensures that each window function integrates to unity
to conserve power.  We pick $\M = \F^{-1}$ without loss of generality, a fact
that we will explain below (see \citet{dillon_et_al2013b} for other contexts where
one might entertain other choices for $\M$, such as taking $\M$ to be
diagonal).

It is important to note that Equation \ref{eq:estimator} estimates power spectra from single-baseline data.  The length of $\x$ is therefore equal to the number of delay bins, and the covariance denoted here relates different delay bins from the same baseline only.  The covariance matrix shown in Figure \ref{fig:cov}, in contrast, includes correlations between delay bins of different baselines as well.  However, its analysis in \S\ref{sec:cov_diag} serves only to  provide a better estimate for $\C$ by taking advantage of redundant baselines.  In other words, our final result in \S\ref{sec:DelayTransApp} will yield a procedure for removing off-diagonal systematics that operates on a per-baseline basis once $\C$ has been estimated, with information from different baselines mixing only in the cross-multiplication to form the power spectrum.

\subsection{Application to Delay-Transformed Visibilities}
\label{sec:DelayTransApp}
We now turn to applying the quadratic estimator formalism to delay-transformed visibilities.
As derived in P12b, $\tau$-modes are a good approximation to $k_\parallel$, and their
mutual orthogonality makes the simplified analysis in \S\ref{sec:dspec_crossmult} a good
approximation to optimal analysis.  As a result, we 
only apply the full quadratic formalism to leading order in order to safeguard against
the mis-modeling of covariances, which are estimated empirically from a finite dataset, and are therefore
imperfect representations of the true covariances.
In this case, the leading-order optimal estimator reduces to the simple prescription given
in Equation \ref{eq:pspec_cosmo},
provided that we used modified visibilities (as shown below) in place of $\tilde{V}(\tau,t)$.

Working in delay space, the $\Q$ matrices become particularly simple, since
delay modes are to an excellent approximation the same as Fourier modes along
the line-of-sight (P12b).  In this approximation,
$\C_{\textrm{sky}}$ is diagonal, with elements equal to the bandpowers, i.e.
\begin{equation}
\label{eq:Csky}
\C^{\textrm{sky}}_{AB} = \delta_{AB} p_A
%\C_{\textrm{sky}} = \left( \begin{array}{ccccccccc}
%p_1 &  0 & 0 & 0 & 0 & & 0 \\
%0 &  p_2 & 0 & 0 & 0 & & 0 \\
%0 &  0 & p_3 & 0 & 0 & \cdots & 0 \\
%0 &  0 & 0 & p_4 & 0 & & 0 \\
%0 &  0 & 0 & 0 & p_5 & & 0 \\
% &  & \vdots &  &  & \ddots  & 0\\
%0 &  0 & 0 & 0 & 0 &0 & p_N \\
%\end{array} \right).
\end{equation}
where $A,B$ index delay modes, and $\delta_{AB}$ is the Kronecker delta function.
It follows that in this basis, the $\Q$ matrices simply pick out the correct delay/Fourier component:
\begin{equation}
\label{eq:Qform}
\Q^\beta_{AB} = \frac{\Omega B}{X^2Y}\left(\frac{2k_{\rm B}}{\lambda^2}\right)^2 \delta_{AB} \delta_{A\beta}.
\end{equation}
Like the sky signal, the instrumental noise covariance can be modeled as a diagonal matrix, so that
\begin{equation}
\label{eq:NoiseCovar}
\N_{AB} = \delta_{AB} \sigma_A^2,
\end{equation}
where $\sigma_A$ is the RMS noise fluctuation in the $A$th delay bin.

With ideal data, where $\C_\textrm{junk} = 0$, the preceding discussion implies
that $\C$ should be diagonal.  However, as we can see from Figure \ref{fig:cov},
the actual data covariance contains non-ideal
off-diagonal entries.  We model the real covariance matrix as
\begin{equation}
\C = \left( \begin{array}{ccccccccc}
r_1^2 &  \varepsilon_{12} r_1 r_2 & \varepsilon_{13} r_1 r_3  &   \\
\varepsilon_{21} r_2 r_1 &  r_2^2 & \varepsilon_{23} r_2 r_3  & \cdots  \\
\varepsilon_{31} r_3 r_1 &  \varepsilon_{32} r_3 r_2 & r_3^2  & \\
 &  \vdots & &   \ddots  \\
\end{array} \right),
\end{equation}
where $r_A \equiv \sqrt{p_A + \sigma_A^2}$ is the total RMS power (sky plus
noise) in the $A$th delay bin, and $\varepsilon_{AB}$ is the coupling constant
between the $A$th and $B$th bins.  Since $\C$ is Hermitian, we note
that $\varepsilon_{AB} = \varepsilon_{BA}^*$.  Note that so far, our
parameterization of $\C$ is still completely general because we have yet to
specify the $\varepsilon$ parameters; the fact that the off-diagonal entries are
scaled to the diagonals is simply a convenient convention driven by the observation
that our empirically derived covariances seem to roughly follow this form.  By comparing
equations \ref{eq:Qdef}, \ref{eq:Csky}, and \ref{eq:NoiseCovar}, we see
that\footnote{It is important to emphasize that in general, systematic contaminants need not be limited to off-diagonal entries.  However, diagonal systematics are difficult to distinguish from signal without other distinguishing features, and so here we set $\C_{\textrm{junk}}$ to be traceless by definition.  In effect, any systematics are included as additional signal along the diagonals of $\C_{\textrm{sky}}$ or $\N$.  An examination of our result will reveal that this amounts to a conservative approach where we leave diagonal systematics untouched in an effort to avoid subtracting cosmological signal.}
\begin{equation}
\C_{\textrm{junk}} = \left( \begin{array}{ccccccccc}
0 &  \varepsilon_{12} r_1 r_2 & \varepsilon_{13} r_1 r_3 &    \\
\varepsilon_{21} r_2 r_1 &  0 & \varepsilon_{23} r_2 r_3 & \cdots   \\
\varepsilon_{31} r_3 r_1 &  \varepsilon_{32} r_3 r_2 & 0 & \\
 &  \vdots & &  \ddots  \\
%0 &  0 & 0 & 0 & 0 &0 & 0 \\
\end{array} \right).
\end{equation}

We will now proceed to compute the various quantities required for our power
spectrum estimator, equation \ref{eq:estimator}.  First, we compute
$\C^{-1}$, which will be our primary tool for mitigating the influence of the
junk contribution.  Instead of performing a brute force inversion of $\C$, we
can make the observation that it is equal to\footnote{This is a special case of
the binomial inverse theorem.}
\begin{equation}
\C^{-1} \equiv (\N  +  \C^{\textrm{sky}}+ \C^{\textrm{junk}})^{-1} = (\N  +  \C^{\textrm{sky}})^{-1} - (\N  +  \C^{\textrm{sky}}+ \C^{\textrm{junk}})^{-1} \C^{\textrm{junk}} (\N  +  \C^{\textrm{sky}})^{-1}.
\end{equation}
Taking advantage of the fact that $(\N  +  \C^{\textrm{sky}}+
\C^{\textrm{junk}})^{-1}$ appears on the right hand side, we can iterate this
expression into a series:
\begin{eqnarray}
\label{eq:series}
\C^{-1} =&& \!\! (\N  +  \C^{\textrm{sky}})^{-1}  - (\N  +  \C^{\textrm{sky}})^{-1} \C^{\textrm{junk}} (\N  +  \C^{\textrm{sky}})^{-1}  \nonumber \\
&&+ (\N  +  \C^{\textrm{sky}})^{-1} \C^{\textrm{junk}} (\N  +  \C^{\textrm{sky}})^{-1} \C^{\textrm{junk}} (\N  +  \C^{\textrm{sky}})^{-1} + \dots
\end{eqnarray}
Writing $\C^{-1}$ in this form makes our results more robust to ``noise" that
arises from our empirical determination of the covariance matrix.  For example,
our series expansion does not require the inversion of $\C_\textrm{junk}$,
which may be unstable under inversion thanks to our imperfect estimation of the
covariance.  One only needs to invert the combination $\N  +
\C^{\textrm{sky}}$, which is trivial because both $\N$ and $\C^\textrm{sky}$
are diagonal.  With a systematic series expansion, we can also henceforth work
only to leading order in the $\varepsilon$ parameters.  We make this choice
based on the intuition that uncertainties in our covariance estimation mean
that higher order terms cannot be reliably estimated.  A leading-order
approximation may therefore be preferable to using an overly ``noisy"
covariance to determine $\C^{-1}$, as that would result in an overfitting of
the noise in the data \citep{tegmark_et_al1998}.

Taking only the first two terms of equation \ref{eq:series}, one obtains
\begin{equation}
\label{eq:ApproxCinv}
\C^{-1} \approx \left( \begin{array}{ccccccccc}
%r_1^{-2} &  -\varepsilon_{12} r^{-1}_1 r^{-1}_2 & -\varepsilon_{13} r^{-1}_1 r^{-1}_3 &   \\
%-\varepsilon_{21} r^{-1}_2 r^{-1}_1 &  r_2^{-2} & -\varepsilon_{23} r^{-1}_2 r^{-1}_3 & \cdots  \\
%-\varepsilon_{31} r^{-1}_3 r^{-1}_1 &  -\varepsilon_{32} r^{-1}_3 r^{-1}_2 & r_3^{-2}      &  \\
1/r_1^{2} &  -\varepsilon_{12}/r_1 r_2 & -\varepsilon_{13}/r_1 r_3 &   \\
-\varepsilon_{21}/r_2 r_1 &  1/r_2^{2} & -\varepsilon_{23}/r_2 r_3 & \cdots  \\
-\varepsilon_{31} r_3 r_1 &  -\varepsilon_{32}/r_3 r_2 & 1/r_3^{2}      &  \\
 &  \vdots & &  \ddots  \\
\end{array} \right).
\end{equation}
Inserting this into equation \ref{eq:FishDef}, it is straightforward to show
that the effects of $\C_\textrm{junk}$ affect the Fisher matrix only to second
order in the $\varepsilon$ parameters and may therefore be neglected:
\begin{equation}
\label{eq:FisherLeading}
F_{\alpha \beta} = \left(\frac{\Omega B}{X^2Y}\right)^2\left(\frac{2k_{\rm B}}{\lambda^2}\right)^4
    \frac{\delta_{\alpha \beta}}{2 r_i^4} + \BigO{\varepsilon^2}.
\end{equation}
The final ingredient that we require before we can estimate power spectra is
the normalization matrix $\M$.  As we remarked above, we will pick $\M =
\F^{-1}$.  If we had instead picked some popular alternatives such as taking
$\M$ to be diagonal, it would have made no difference, since we have just shown
that the Fisher matrix is diagonal to leading order.

With a diagonal Fisher matrix, we can further simplify our power spectrum
estimator.  We first define a family of matrices
\begin{equation}
\G_\alpha \equiv \frac{1}{2} \sum_\beta M_{\alpha \beta} \Q^\beta
\end{equation}
so that our estimator can be written compactly as
\begin{equation}
\label{eq:CompactEst}
\widehat{p}_\alpha = | \G_\alpha^{1/2} \C^{-1} \x |^2.
\end{equation}
Inserting $\M = \F^{-1}$ as well as equations \ref{eq:Qform} and 
\ref{eq:FisherLeading} into our definition of $\G_\alpha$ yields
\begin{equation}
\G^\alpha_{AB} = \frac{X^2Y}{\Omega B}\left(\frac{\lambda^2}{2k_{\rm B}}\right)^2
    r_\alpha^4 \delta_{AB} \delta_{A\alpha}.
\end{equation}
Substituting this and equation \ref{eq:ApproxCinv} into equation \ref{eq:CompactEst}, we obtain
\begin{equation}
\label{eq:finalAutoEst}
\widehat{p}_\alpha = \frac{X^2Y}{\Omega B}\left(\frac{\lambda^2}{2k_{\rm B}}\right)^2 \Bigg{|} x_\alpha - \sum_{\beta \neq \alpha} \varepsilon_{\alpha \beta} \frac{r_\alpha}{r_\beta} x_\beta \Bigg{|}^2.
\end{equation}
This, then, is our leading-order optimal estimator for the bandpower $p_\alpha$
in the presence of a ``junk" covariance that couples different delay bins.
Intuitively, our estimator instructs us to subtract from each delay bin a
predicted leakage from all other bins, which is determined by a coupling
constant and the ratio of RMS power levels between the bins\footnote{If robust models of diagonal systematics were available, their inclusion in $\C_\textrm{junk}$ would simply result in the elimination of the $\beta \neq \alpha$ restriction on the sum in Equation \ref{eq:finalAutoEst}.}.  The resulting
quantity is then squared to yield a final power estimate.

Written in this way, it is also clear how one would generalize the result to
cross-power spectra.  The optimal estimator for cross-power spectra between
data from baseline $i$ and baseline $j$, for example, would be
\begin{equation}
\widehat{p}_\alpha = \frac{X^2Y}{\Omega B}\left(\frac{\lambda^2}{2k_{\rm B}}\right)^2 \left( x_\alpha - \sum_{\beta \neq \alpha} \varepsilon_{\alpha \beta} \frac{r_\alpha}{r_\beta} x_\beta \right)_i^\dagger 
\left( x_\alpha - \sum_{\gamma \neq \alpha} \varepsilon_{\alpha \gamma} \frac{r_\alpha}{r_\gamma} x_\gamma \right)_j.
\label{eq:cov_diag}
\end{equation}

In summary, if we define
\begin{equation}
\Vt^\prime_i(\tau_\alpha)\equiv \left(x_\alpha - \sum_{\beta \neq \alpha} \varepsilon_{\alpha \beta} \frac{r_\alpha}{r_\beta} x_\beta\right)_i,
\label{eq:vtildeprime}
\end{equation}
then Equation \ref{eq:cov_diag} reveals that Equation \ref{eq:pspec_cosmo} is to leading order an
optimal estimator, provided we use $\Vt^\prime$ in lieu of $\Vt$.

\bibliographystyle{hapj}
\bibliography{biblio}

\end{document}